\documentclass[10pt,twocolumn,letterpaper]{article}

\usepackage{wacv}
\usepackage{times}
\usepackage{epsfig}
\usepackage{graphicx}
\usepackage{amsmath}
\usepackage{amssymb}

\makeatletter
\@namedef{ver@everyshi.sty}{}
\makeatother
\usepackage{tikz}
\usepackage{inset}
\usepackage{mwe,subcaption}
\usetikzlibrary{spy,calc,shapes,arrows,positioning, chains}
\tikzset{boximg/.style={remember picture,red,thick,draw,inner sep=0pt,outer sep=0pt}}
\usepackage{caption}

\usepackage[accsupp]{axessibility}  

\def\wacvPaperID{7} 

\wacvfinalcopy 

\ifwacvfinal
\usepackage[breaklinks=true,bookmarks=false]{hyperref}
\else
\usepackage[pagebackref=true,breaklinks=true,colorlinks,bookmarks=false]{hyperref}
\fi

\begin{document}

\title{Joint Multi-Scale Tone Mapping and Denoising for HDR Image Enhancement}

\author{
Litao Hu\\
Purdue University\\
{\tt\small hu430@purdue.edu}
\and
Huaijin Chen\\
Sensebrain Technology\\
{\tt\small chenhuaijin@sensebrain.ai}
\and
Jan P. Allebach\\
Purdue University\\
{\tt\small allebach@purdue.edu}
}

\maketitle
\thispagestyle{empty}

\begin{abstract}

  An image processing unit (IPU), or image signal processor (ISP) for high dynamic range (HDR) imaging usually consists of demosaicing, white balancing, lens shading correction, color correction, denoising, and tone-mapping. Besides noise from the imaging sensors, almost every step in the ISP introduces or amplifies noise in different ways, and denoising operators are designed to reduce the noise from these sources. Designed for dynamic range compressing, tone-mapping operators in an ISP can significantly amplify the noise level, especially for images captured in low-light conditions, making denoising very difficult. Therefore, we propose a joint multi-scale denoising and tone-mapping framework that is designed with both operations in mind for HDR images. Our joint network is trained in an end-to-end format that optimizes both operators together, to prevent the tone-mapping operator from overwhelming the denoising operator. Our model outperforms existing HDR denoising and tone-mapping operators both quantitatively and qualitatively on most of our benchmarking datasets. Code available at: \href{https://github.com/hulitaotom/Joint-Multi-Scale-Tone-Mapping-and-Denoising-for-HDR-Image-Enhancement}{Joint-Multi-Scale-Tone-Mapping-and-Denoising-for-HDR-Image-Enhancement}.
\end{abstract}

\section{Introduction}

HDR image enhancement is a complex task consisting of two major sub-tasks, namely image tone mapping and image denoising. In a typical ISP for HDR images, tone-mapping operators (TMOs) and image denoising operators are separate components. A tone-mapping operator aims to compress the dynamic range of the HDR image by brightening the low-lights and dimming the high-lights so that the HDR content can be viewed on low-dynamic-range displays preserving its perceptual contrast. Many HDR images are captured under low-exposure settings or low-light conditions, to capture and preserve a wider dynamic range without over-exposing the highlight regions. As a result, noise in HDR images can be a very challenging problem for any existing denoising algorithms designed for standard-dynamic-range (SDR) images. 
Meanwhile, denoising often requires the inputs to be in a linear color space, which can hardly hold true due to the multi-exposure HDR capture and non-linear tone-mapping. 

With the denoising and tone-mapping treated as separate modules in conventional ISPs, it is natural for us to wonder which operator should be applied first. Denoising after TMOs \cite{noise_estimation} can pose significant challenges to denoising, since TMOs are usually highly non-linear and tends to amplify the noise. Denoising before TMOs, in contrast, may cause the remaining noise after denoising being significantly amplified by the subsequent TMOs, requiring extra denoising steps. Therefore, it makes sense for us to design the denoising and tone-mapping jointly.
Meanwhile, we observe that both denoising and TMOs can benefit from multi-scale processing. For instance, tone-mapping can be achieved by adjusting the brightness to the high- and low-frequency component of the HDR image differently \cite{durand2002fast}, and denoising on decomposed multi-scale inputs also receives better results than the single-scale counterpart \cite{multiscale_DCT}.

Thus, in this paper, we propose to \textit{jointly} optimize the two operators within a \textit{multi-scale} network for HDR image enhancement and showcase that our proposed framework achieves better results than state-of-the-art HDR tone-mapping methods and separately performed denoising and tone-mapping procedures on recent HDR benchmark datasets both quantitatively and qualitatively.
Our key contributions are the following:
\begin{itemize}
    \item We propose a novel learned differentiable single image denoising module based on discrete cosine transform (DCT);
    \item We propose a multi-scale image enhancement framework that jointly optimizes the TMOs and denoising operators for HDR image enhancement;
    \item We investigate the effect of different orderings of the TMOs and denoising operators and demonstrate that having TMOs applied first yields better final results.
\end{itemize}

\section{Related Work}


\subsection{HDR Image Tone Mapping}

Tone-mapping algorithms can be categorized into learning-based methods and traditional methods depending on whether training is needed. With the advancement in deep learning, learning-based TMOs are become more advantageous, especially in challenging cases. Within the learning-based category, methods can be further divided into supervised \cite{Chen_2018_CVPR, Chen2018Retinex, LI201815, Wang_2019_CVPR, Chen_2019_ICCV, Jiang_2019_ICCV, KinD, Ren, Wang, Xu_2020_CVPR, Fan, Zhu_Pan_Chen_Yang_2020, SIDGAN, Lv, DSLR, hu2020relation, Hu2020DocumentIQ}, semi-supervised \cite{Yang_2020_CVPR}, and unsupervised \cite{ExCNet, EnlightenGAN, Guo_2020_CVPR, RRDNet, DLN, TBEFN, hu2021deep}, depending on whether ground-truth is provided during training. In general, most supervised TMOs outperform semi-supervised and unsupervised TMOs, while the latter requires fewer paired training samples and thus can be easily implemented in a wide range of applications.

In particular, multi-scale structures such as the Laplacian pyramid has been used by many learning-based image enhancement algorithms including \cite{10.5555/2969239.2969405}, \cite{8100101}, and \cite{afifi2021learning}.

In our work, we adopt the CSRNet \cite{he2020conditional} as our backbone tone-mapping network for its decent performance, lightweight, and fully convolutional structure that makes it compatible with any size input.

\subsection{HDR Image Denoising}

Deep learning has been applied extensively to the image denoising field \cite{denoise_survey}, outperforming many of the traditional denoising algorithms. Based on the number of images used as input for a given scene, these deep learning-based image denoising methods can be categorized into single-image denoising and multi-image denoising. Single-image denoising \cite{ducnn, ffdnet, Guo2019Cbdnet, Chen_gan, pmlr-v80-lehtinen18a, Liu_2020_CVPR_Workshops, Bao_2020_CVPR_Workshops, GradNet, Vaksman_2020_CVPR_Workshops, Anwar_2020_CVPR_Workshops, hu2019frame, Hu2019NonnativeCD} aims to denoise a target scene with only a single shot from the scene. On the other hand, multi-image denoising \cite{burst_denoise} aims to denoise a target scene by utilizing multiple shots from the same scene. A common strategy in multi-image denoising utilizes images from a shot burst, providing a richer spatial-temporal context for the model. However, this requires image registration preprocessors for aligning and merging multiple images from a shot burst, which can cause other artifacts and is usually computationally intensive. 

Based on whether a multi-scale structure is used, denoising methods can also be classified into single-scale methods and multi-scale methods. Multi-scale denoising has been an effective strategy for extracting and providing multi-scale information to the denoising models \cite{Liu_2020_CVPR_Workshops, Bao_2020_CVPR_Workshops, Vaksman_2020_CVPR_Workshops}.

In our work, we adopt a single-image multi-scale denoising framework and propose a novel deep DCT denoising module inspired by the traditional DCT denoising algorithm.

\subsection{Joint Tone Mapping and Denoising for HDR Images}

For a noisy raw HDR image, different modules in an ISP will have different effects on the noise distribution. Modules such as demosaicing and tone-mapping can sometimes boost the noise significantly if not designed wisely. Therefore, some recent works try to combine denoising with these modules, to get a better measurement of the noise distribution and provide more intermediate information to the denoising algorithm. \cite{joint_dm, filippos, thibaud, Liu_2020_CVPR} proposed to jointly perform image demosaicing and image denoising. However, a major drawback of these methods is that since demosaicing is usually performed at the beginning of an ISP, the remaining noise after the joint denoising can still be amplified by the following operations, especially those highly non-linear ones such as tone mapping. As a result, extra denoising modules are often needed to clean up the noise remaining from these operations. A joint tone mapping and denoising framework, on the other hand, does not have such a drawback since TMOs are usually placed near the end of an ISP pipeline. Moreover, by jointly optimizing two modules, the TMO can provide information and context on where the noise gets amplified more. \cite{noise_estimation} has proved the effectiveness of applying joint tone mapping and denoising on HDR contents.

In our work, we extend the idea of joint tone mapping and denoising to a multi-scale framework, where the TMOs and denoising operators are jointly optimized at multiple scales in a Laplacian pyramid.


\begin{figure*}[ht!]
\begin{center}
  \includegraphics[width=\linewidth]{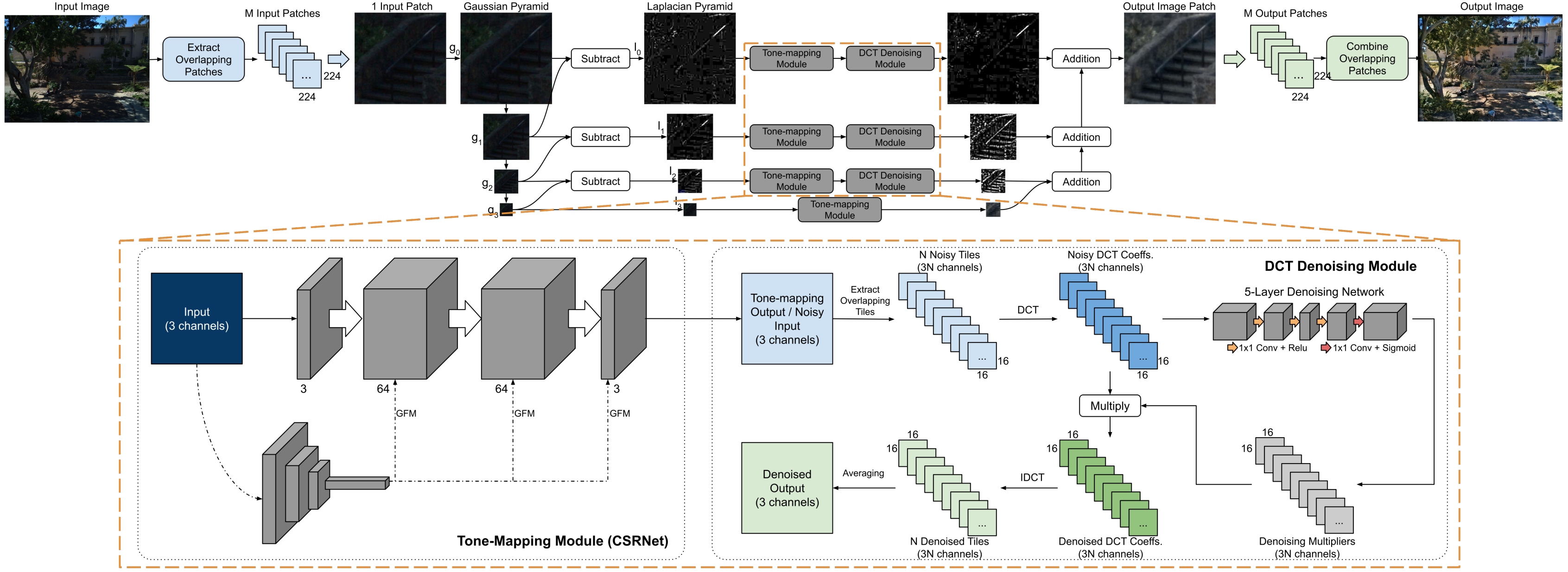}
\end{center}
  \caption{Our proposed HDR image enhancement framework. The upper portion of the figure shows a high-level workflow of our framework and the multi-scale structure. The lower portion of the figure shows more details of the tone-mapping module and the denoising module.}
\label{fig:overall}
\end{figure*}

\section{Proposed Method}

In this section, we will present our joint multi-scale tone mapping and denoising framework for an HDR image in detail. The overall structure of our proposed framework is shown in Figure \ref{fig:overall}.


\subsection{Deep DCT Denoising for a Single HDR Image}

We will first introduce our deep DCT denoising network designed for single HDR image denoising. The architecture of our deep DCT denoising algorithm for a single HDR image is shown in Figure \ref{fig:overall}.

DCT denoising was first proposed in 2011 by \cite{DCT_denoise} and was widely used because of its simplicity and effectiveness. Later in 2017, \cite{multiscale_DCT} extended the DCT denoising algorithm to a multi-scale setting. The deep DCT denoising network we propose is a learnable DCT denoising algorithm that combines the simplicity of the traditional DCT denoising algorithm and the effectiveness of modern deep neural networks.

In traditional DCT denoising algorithms, a preset threshold is chosen to zero out DCT coefficients below the threshold in order to create sparsity in the DCT domain of local image tiles. However, this operation is not differentiable and thus is not suitable for our learning-based framework. Therefore, instead of applying a learned threshold to the DCT coefficients, we multiply the DCT coefficients with learned re-weighting maps with re-weighting values between 0 and 1. In a re-weighting map, a value close to 0 is equivalent to the case where the corresponding DCT coefficient is below the threshold, while a value close to 1 is equivalent to the case where the DCT coefficient is above the threshold. Using re-weighting maps not only allows for differentiable parameters but also provides more flexibility to the denoising module, leading to better denoising performance.

The input to our DCT denoising module is a 3-channel noisy image patch in RGB. As the first step in the denoising module, the input image patch will be split into overlapping tiles in size of $16\times 16$, which will then be reshaped into a stack of tiles. We adopt $16\times 16$ as the size of all local image tiles following \cite{DCT_denoise}. Organizing the tiles this way allows us to perform the DCT easily for every tile. These noisy DCT tiles will then be passed to a 5-layer CNN to estimate a denoising multiplier within the DCT domain, which will then be multiplied element-by-element by the noisy DCT tiles to get the same number of denoised DCT tiles. These denoised DCT tiles will then go through an inverse discrete cosine transform (IDCT), as well as a tile-averaging operation, to reconstruct the denoised output image patch. Note that the tile-averaging operation here is a simple averaging operation applied to the overlapping pixels among neighboring tiles. For example, if a pixel in the output image patch is overlapped by 4 neighboring tiles that reconstruct this output image patch, the value for this pixel will then be the average of the 4 values of the corresponding locations in these 4 overlapping tiles.

Our DCT denoising process can be represented by Equation \ref{eq:denoising}, where $C_{noisy}$ and $C_{denoised}$ represent the DCT coefficients of the noisy input and the denoised output, respectively. $\mathcal{F}$ represents the non-linear function learned by the 5-layer CNN for estimating the denoising multipliers, and $\odot$ represents element-by-element multiplication. Note that the 5-layer CNN here for estimating the denoising multipliers contains only $1\times 1$ 2D convolutional layers. We made this design choice intentionally because we believe that in the DCT domain, the coefficients at different frequencies should be considered separately, while coefficients at the same frequency should be grouped together. Thus, convolutional kernels that are commonly used in the image domain, such as a $3 \times 3$ 2D convolutional layer, would not be suitable here.
\begin{align}
    C_{denoised} = C_{noisy} \odot \mathcal{F}(C_{noisy})
    \label{eq:denoising}
\end{align}

\subsection{Joint Multi-Scale Tone-mapping and Denoising Framework}

The other important component of our proposed framework is its joint multi-scale tone-mapping and denoising framework. The overall structure of our framework and a closer look at the multi-scale structure are shown in Figure \ref{fig:overall}.


Multi-scale structures have been used extensively in many image processing and computer vision tasks for their effectiveness in producing finer results than their single-scale counterparts. Some \cite{Bao_2020_CVPR_Workshops, Vaksman_2020_CVPR_Workshops} incorporate the multi-scale aspect by simply downsampling the input a few times, and then processing each scale separately, while others \cite{Liu_2020_CVPR_Workshops, multiscale_DCT} utilize domain transformations such as the discrete wavelet transform (DWT) and DCT.

In our proposed framework, we utilize the Laplacian pyramid to obtain multiple scales of an input image. To be specific, a noisy input RGB image will first be divided into half-overlapping image patches of size $224 \times 224$ pixels. Then every image patch will be decomposed into four Gaussian pyramid layers ($g_0$, $g_1$, $g_2$, $g_3$), From there, three Laplacian layers ($l_0$, $l_1$, $l_2$) will be computed from the neighboring Gaussian layers, which then forms a four-layer Laplacian pyramid along with the Gaussian base layer ($g_3$ or $l_3$).

From the Laplacian pyramid, the Laplacian layers $l_0$, $l_1$, and $l_2$ will be passed to their corresponding tone mapping networks, followed by their corresponding deep DCT denoising modules, while the base layer ($g_3$ or $l_3$) will only be tone-mapped. We don't denoise the base layer because most high-frequency noise will be separated to the upper layers and it is unnecessary for a denoising operator in the base layer. This can help reduce the model size and speed up inference. 

These tone-mapped and denoised output Laplacian layers will then be combined to reconstruct the corresponding tone-mapped and denoised output RGB image patch, following the inverse process of obtaining the Laplacian pyramid. Finally, all the output image tiles will be combined into a single output RGB image using a modified raised-cosine filter, following the same way of merging overlapping image tiles in \cite{rcf}. As shown in our results, this modified raised-cosine filter does a very good job merging image tiles without showing any obvious artifacts.

\subsection{Training Scheme and Loss Functions}

To better optimize both the tone-mapping networks and the denoising modules, we adopt a three-phase training scheme in our experiments. To be specific, we train the model in three consecutive phases, namely tone-mapping training, denoising training, and joint training. 

In each training phase, a different part of the joint framework will be trained and different loss functions will be used to serve each training purpose. 

The first training phase, i.e. the tone-mapping training phase, focuses on pre-training the tone-mapping networks. In this phase, the denoising modules will be temporarily moved out of the joint framework, meaning that each layer in the pyramid will only go through a tone-mapping network. The loss function for this phase is computed according to Equations \ref{eq:loss1-1} and \ref{eq:loss1-2}, where $L_{l_0}$, $L_{l_1}$, $L_{l_2}$, $L_{l_3}$ are the L1 loss computed at each Laplacian layer, and $l_i^{(1)}$ and $l_i$ represent, respectively, the tone-mapped Laplacian layer and the corresponding Laplacian layer from the tone-mapped ground-truth image. Here, $\lambda_0$, $\lambda_1$, $\lambda_2$, and $\lambda_3$ are hyper-parameters, and the selected values for them can be found in Section 4.1. 
\begin{align}
    L_{p1} &= \lambda_0 L_{l_0}^{(1)} + \lambda_1 L_{l_1}^{(1)} + \lambda_2 L_{l_2}^{(1)} + \lambda_3 L_{l_3}^{(1)} \label{eq:loss1-1}\\
    L_{l_i}^{(1)} &= |l_i^{(1)} - l_i|,\ i=0,1,2,3
    \label{eq:loss1-2}
\end{align}

The second phase, i.e. the denoising phase, focuses on pre-training the denoising networks. The denoising modules are added back to the joint framework during this phase, while the pre-trained tone-mapping networks are fixed. The loss function for this phase is computed according to Equations \ref{eq:loss2-1}-\ref{eq:loss2-3}, where $L_{l_0}$, $L_{l_1}$, $L_{l_2}$ are the L1 loss computed at each Laplacian except for $l_3$, and $l_i^{(0)}$ and $l_i^{(2)}$ represent, respectively, the input Laplacian layer and the tone-mapped denoised Laplacian layer. $L_{denoise}$ measures the denoising loss by computing the L1 loss between the reconstructed denoised output and the clean ground-truth image without tone-mapping ($Y_{merged}$). Here, we use $R$ to represent the reconstruction operator. Again, $\lambda_0$, $\lambda_1$, $\lambda_2$, and $\lambda_d$ are hyper-parameters, and the selected values for them can be found in Section 4.1. 
\begin{align}
    L_{p2} &= \lambda_0 L_{l_0}^{(2)} + \lambda_1 L_{l_1}^{(2)} + \lambda_2 L_{l_2}^{(2)} + \lambda_d L_{denoise} \label{eq:loss2-1}\\
    L_{l_i}^{(2)} &= |l_i^{(2)} - l_i|,\ i=0,1,2\\
    L_{denoise} &= |R(l_0^{(2)}, l_1^{(2)}, l_2^{(2)}, l_3^{(0)}) - Y_{merged}|
    \label{eq:loss2-3}
\end{align}

The last phase, i.e. the joint training phase, trains both the tone-mapping networks and the denoising modules together. The loss function for this phase is simply an L1 loss computed between the reconstructed output from the tone-mapped denoised Laplacian layers and the clean tone-mapped ground-truth image $Y_{final}$, as shown in Equation \ref{eq:loss3}.
\begin{align}
    L_{p3} &= |R(l_0^{(2)}, l_1^{(2)}, l_2^{(2)}, l_3^{(1)}) - Y_{final}|
    \label{eq:loss3}
\end{align}


\begin{figure*}
\captionsetup[subfigure]{labelformat=empty}
  \centering
    \begin{subfigure}[b]{0.11\textwidth}
    \begin{tikzpicture}[zoomboxarray, zoomboxes below]
        \node [image node] { \includegraphics[width=0.95\textwidth]{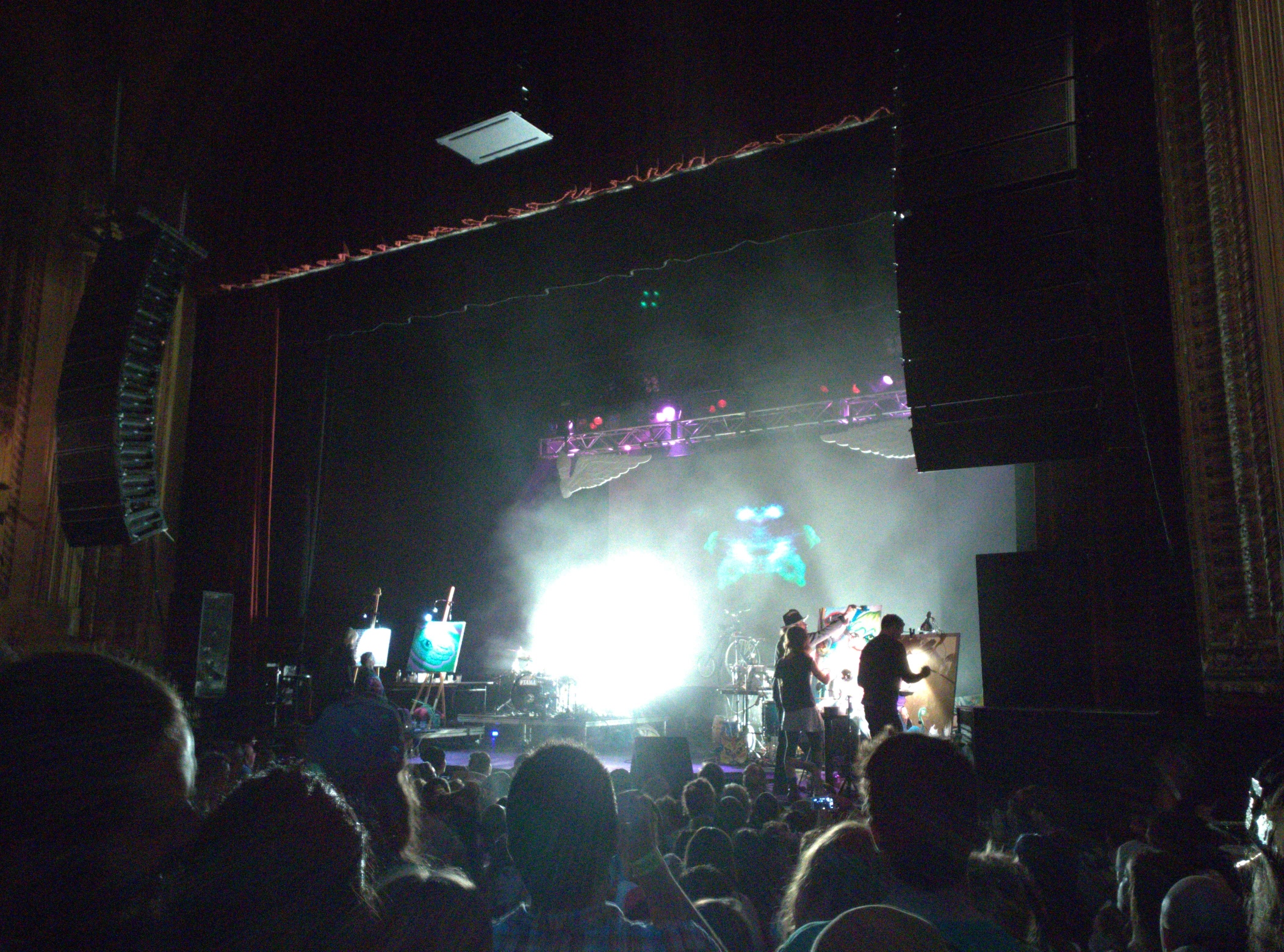} };
        \zoombox[magnification=10, color code=red]{0.175,0.8}
        \zoombox[magnification=10, color code=yellow]{0.61,0.15}
        \zoombox[magnification=10, color code=orange]{0.37,0.29}
        \zoombox[magnification=10,color code=lime]{0.82,0.91}
    \end{tikzpicture}\vspace{-5px}\caption{Input}\vspace{3px} \end{subfigure} \begin{subfigure}[b]{0.11\textwidth}
    \begin{tikzpicture}[zoomboxarray, zoomboxes below]
        \node [image node] { \includegraphics[width=0.95\textwidth]{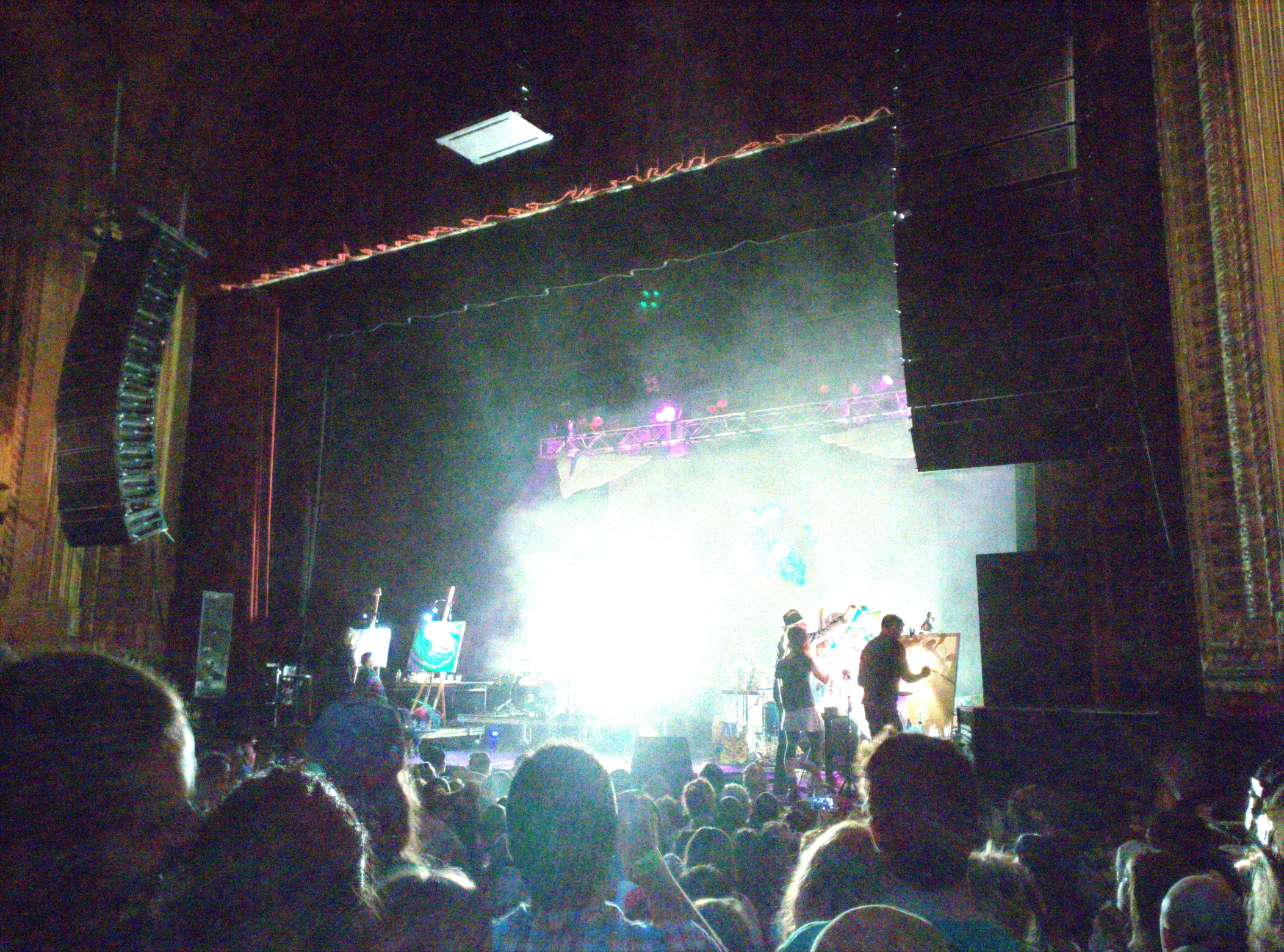} };
        \zoombox[magnification=10, color code=red]{0.175,0.8}
        \zoombox[magnification=10, color code=yellow]{0.61,0.15}
        \zoombox[magnification=10, color code=orange]{0.37,0.29}
        \zoombox[magnification=10,color code=lime]{0.82,0.91}
    \end{tikzpicture}\vspace{-5px}\caption{DSLR}\vspace{3px} \end{subfigure} \begin{subfigure}[b]{0.11\textwidth}
    \begin{tikzpicture}[zoomboxarray, zoomboxes below]
        \node [image node] { \includegraphics[width=0.95\textwidth]{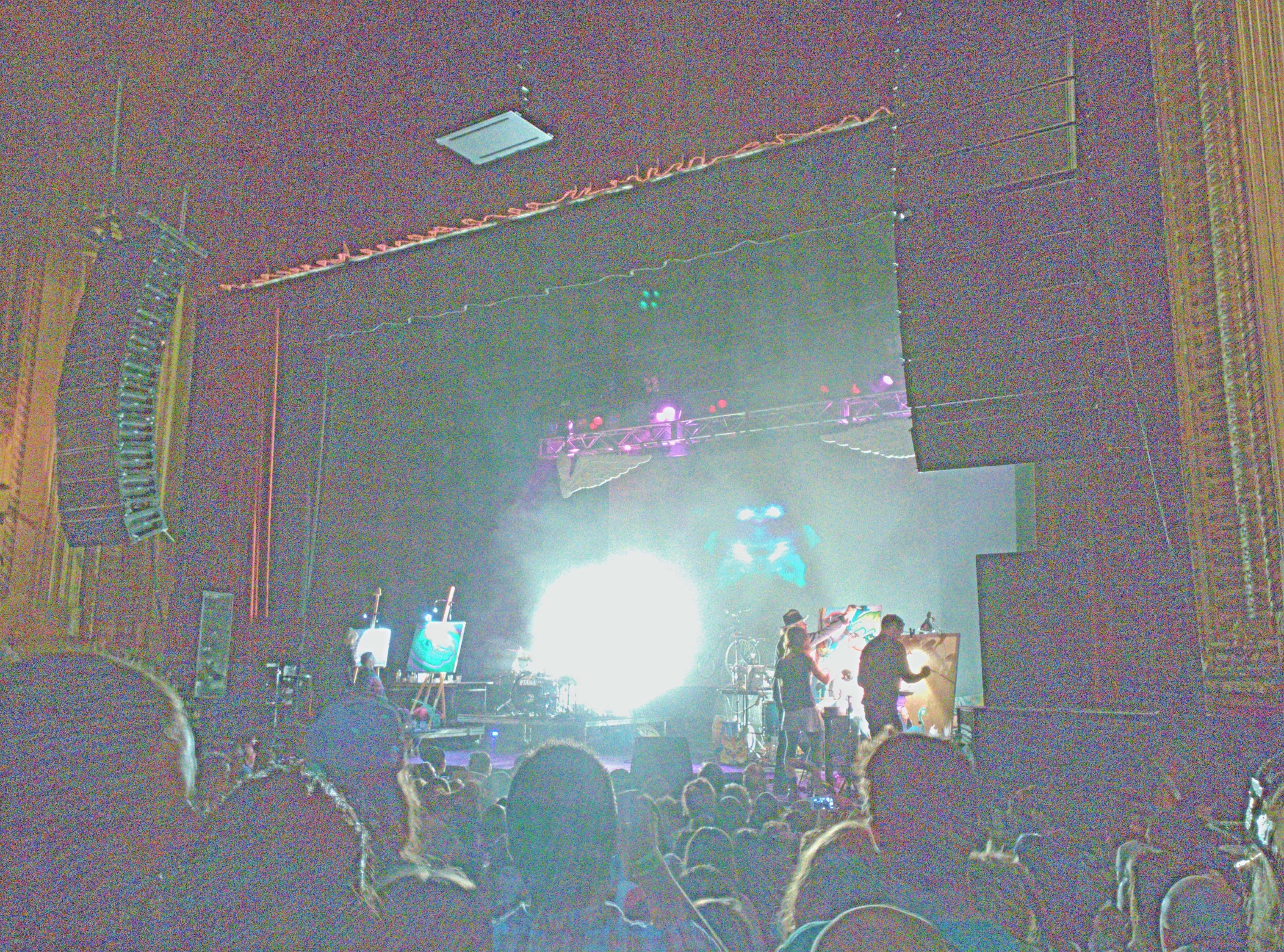} };
        \zoombox[magnification=10, color code=red]{0.175,0.8}
        \zoombox[magnification=10, color code=yellow]{0.61,0.15}
        \zoombox[magnification=10, color code=orange]{0.37,0.29}
        \zoombox[magnification=10,color code=lime]{0.82,0.91}
    \end{tikzpicture}\vspace{-5px}\caption{TBEFN}\vspace{3px} \end{subfigure} \begin{subfigure}[b]{0.11\textwidth}
    \begin{tikzpicture}[zoomboxarray, zoomboxes below]
        \node [image node] { \includegraphics[width=0.95\textwidth]{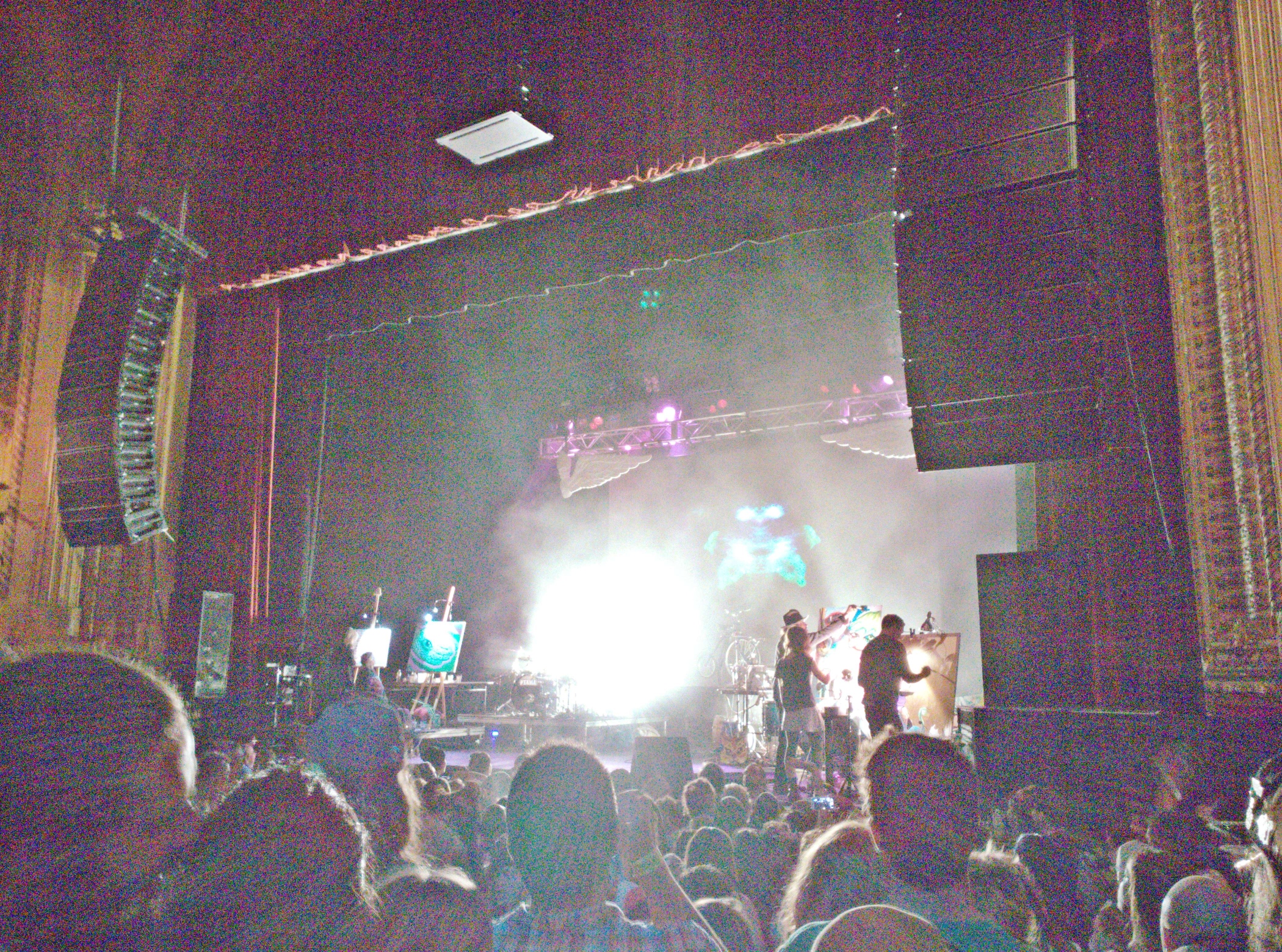} };
        \zoombox[magnification=10, color code=red]{0.175,0.8}
        \zoombox[magnification=10, color code=yellow]{0.61,0.15}
        \zoombox[magnification=10, color code=orange]{0.37,0.29}
        \zoombox[magnification=10,color code=lime]{0.82,0.91}
    \end{tikzpicture}\vspace{-5px}\caption{Zero-DCE++}\vspace{3px} \end{subfigure} \begin{subfigure}[b]{0.11\textwidth}
    \begin{tikzpicture}[zoomboxarray, zoomboxes below]
        \node [image node] { \includegraphics[width=0.95\textwidth]{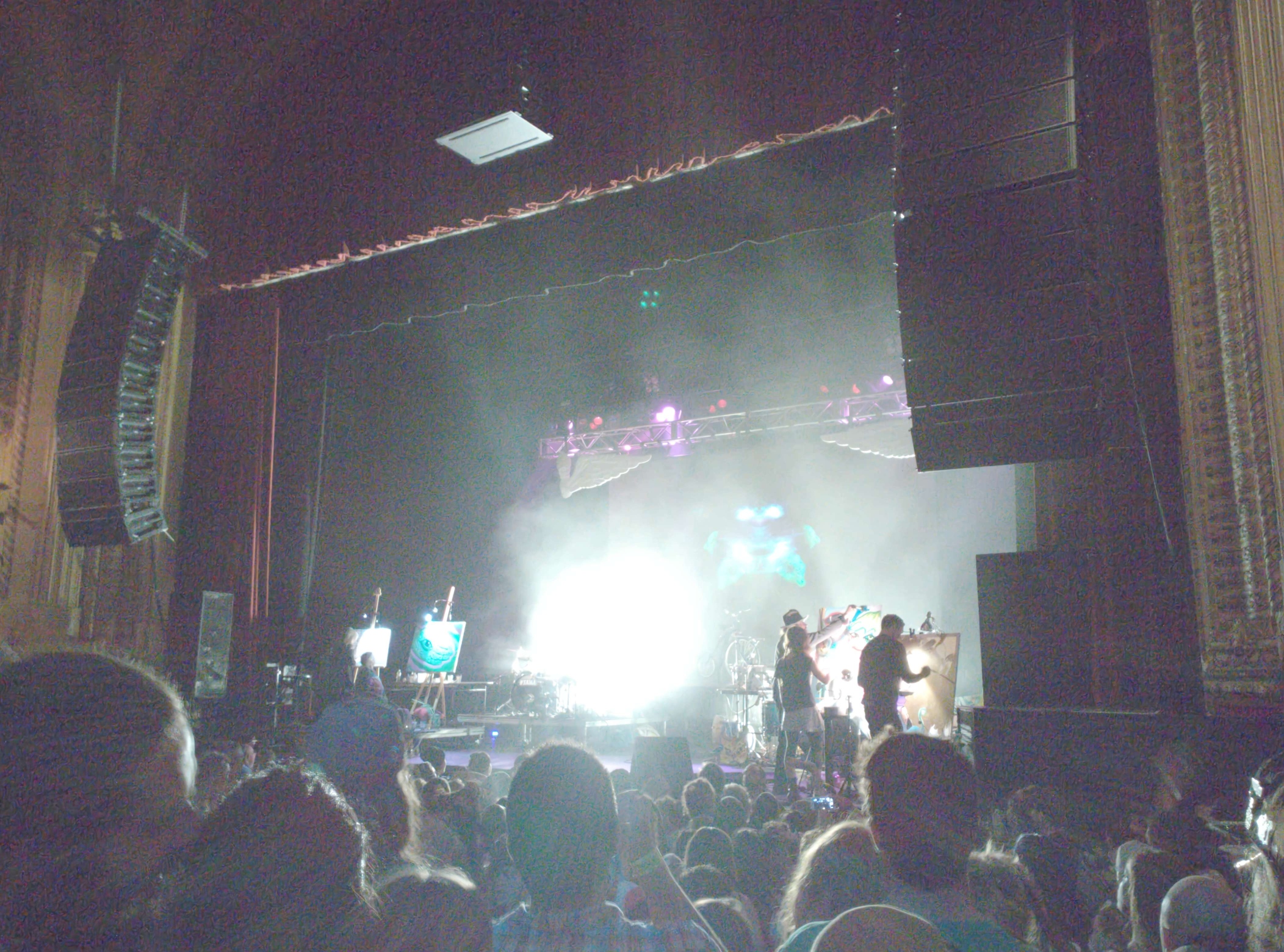} };
        \zoombox[magnification=10, color code=red]{0.175,0.8}
        \zoombox[magnification=10, color code=yellow]{0.61,0.15}
        \zoombox[magnification=10, color code=orange]{0.37,0.29}
        \zoombox[magnification=10,color code=lime]{0.82,0.91}
    \end{tikzpicture}\vspace{-5px}\caption{FFDNet+GC}\vspace{3px} \end{subfigure} \begin{subfigure}[b]{0.11\textwidth}
    \begin{tikzpicture}[zoomboxarray, zoomboxes below]
        \node [image node] { \includegraphics[width=0.95\textwidth]{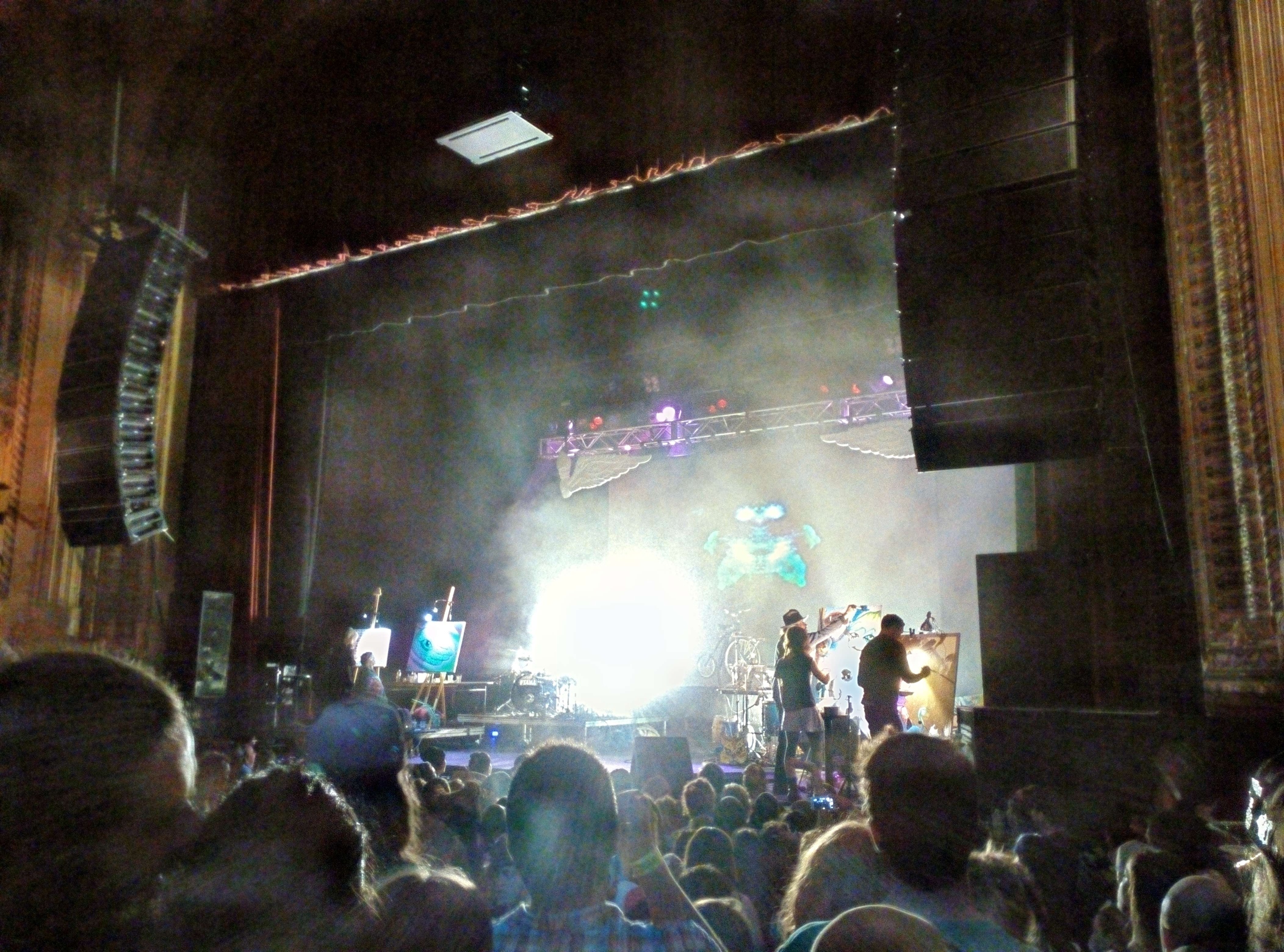} };
        \zoombox[magnification=10, color code=red]{0.175,0.8}
        \zoombox[magnification=10, color code=yellow]{0.61,0.15}
        \zoombox[magnification=10, color code=orange]{0.37,0.29}
        \zoombox[magnification=10,color code=lime]{0.82,0.91}
    \end{tikzpicture}\vspace{-5px}\caption{DFTL (Ours)}\vspace{3px} \end{subfigure} \begin{subfigure}[b]{0.11\textwidth}
    \begin{tikzpicture}[zoomboxarray, zoomboxes below]
        \node [image node] { \includegraphics[width=0.95\textwidth]{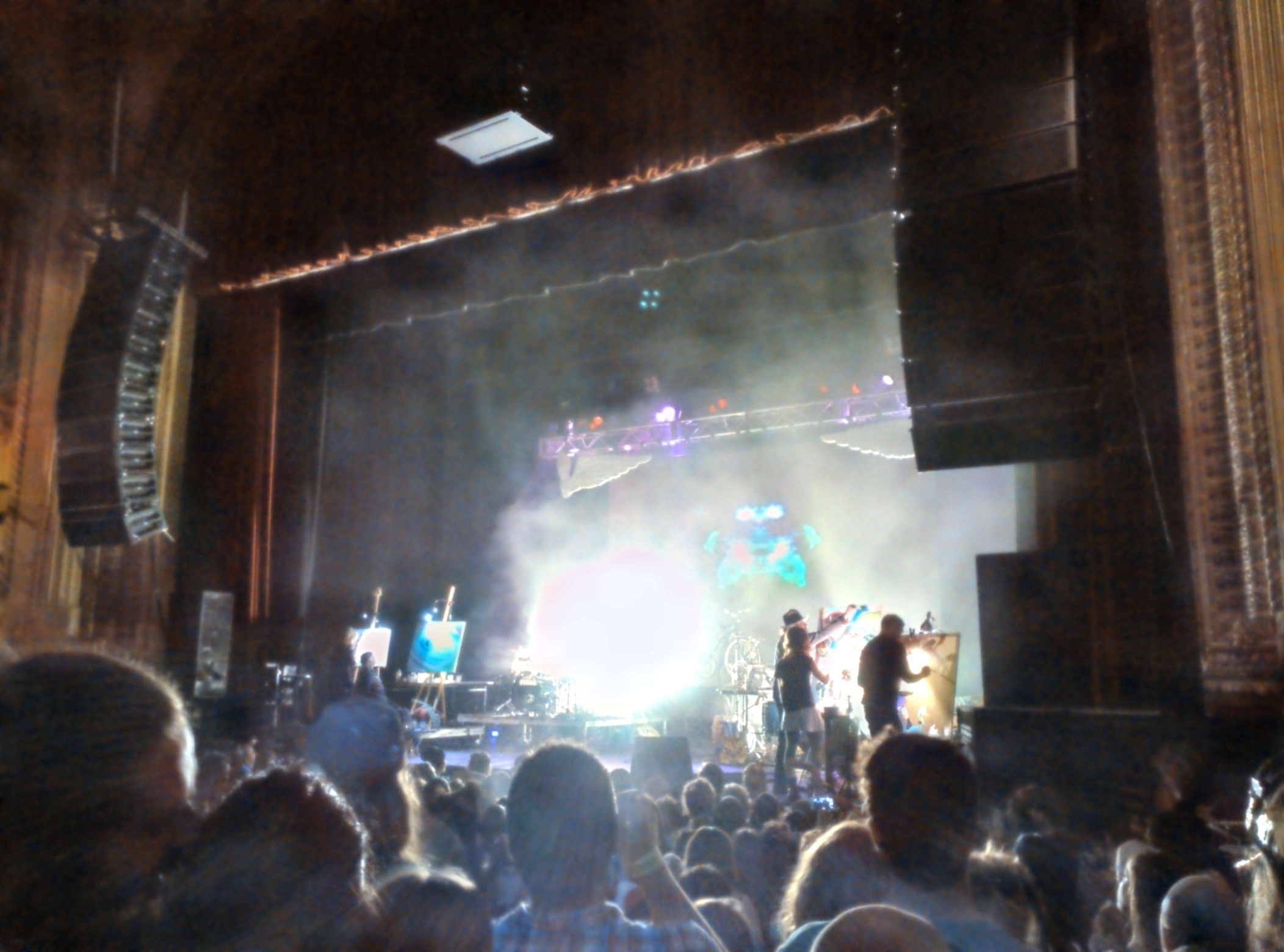} };
        \zoombox[magnification=10, color code=red]{0.175,0.8}
        \zoombox[magnification=10, color code=yellow]{0.61,0.15}
        \zoombox[magnification=10, color code=orange]{0.37,0.29}
        \zoombox[magnification=10,color code=lime]{0.82,0.91}
    \end{tikzpicture}\vspace{-5px}\caption{TFDL (Ours)}\vspace{3px} \end{subfigure} \begin{subfigure}[b]{0.11\textwidth}
    \begin{tikzpicture}[zoomboxarray, zoomboxes below]
        \node [image node] { \includegraphics[width=0.95\textwidth]{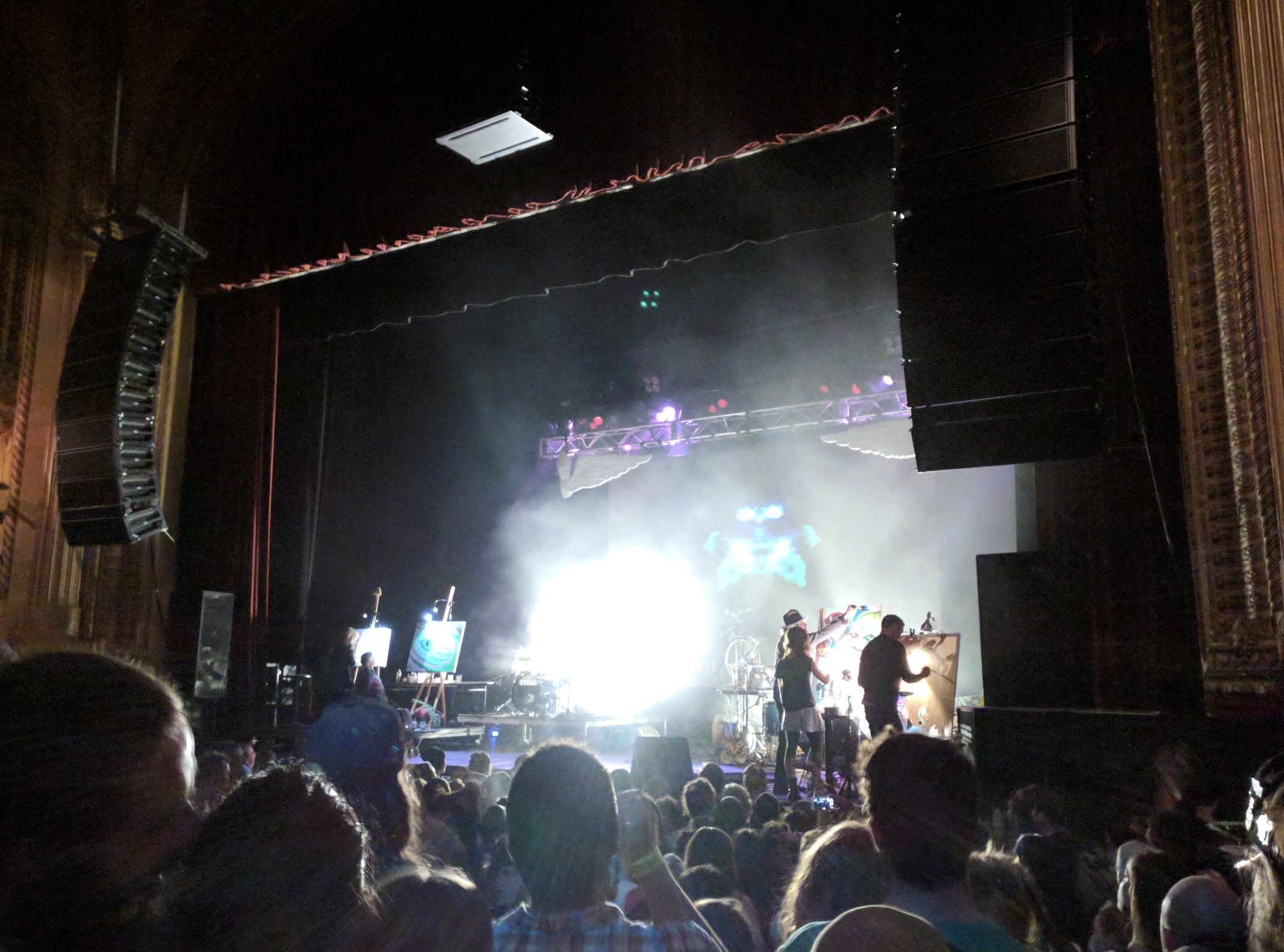} };
        \zoombox[magnification=10, color code=red]{0.175,0.8}
        \zoombox[magnification=10, color code=yellow]{0.61,0.15}
        \zoombox[magnification=10, color code=orange]{0.37,0.29}
        \zoombox[magnification=10,color code=lime]{0.82,0.91}
    \end{tikzpicture}\vspace{-5px}\caption{Ground-truth}\vspace{3px} \end{subfigure} \begin{subfigure}[b]{0.11\textwidth}
    \begin{tikzpicture}[zoomboxarray, zoomboxes below]
        \node [image node] { \includegraphics[width=0.95\textwidth]{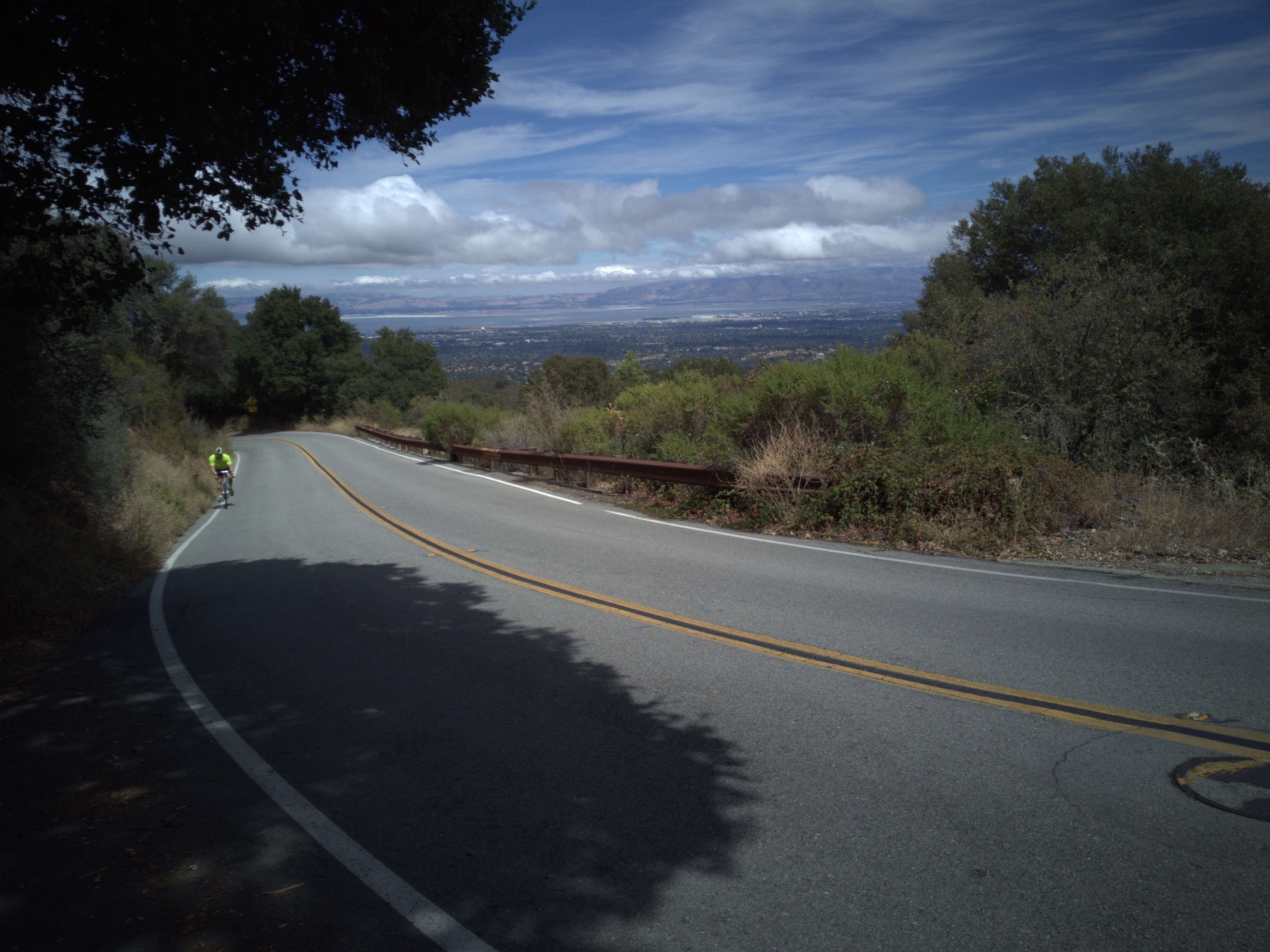} };
        \zoombox[magnification=10, color code=red]{0.175,0.8}
        \zoombox[magnification=10, color code=yellow]{0.51,0.68}
        \zoombox[magnification=10, color code=orange]{0.175,0.51}
        \zoombox[magnification=10,color code=lime]{0.82,0.71}
    \end{tikzpicture}\vspace{-5px}\caption{Input}\vspace{3px} \end{subfigure} \begin{subfigure}[b]{0.11\textwidth}
    \begin{tikzpicture}[zoomboxarray, zoomboxes below]
        \node [image node] { \includegraphics[width=0.95\textwidth]{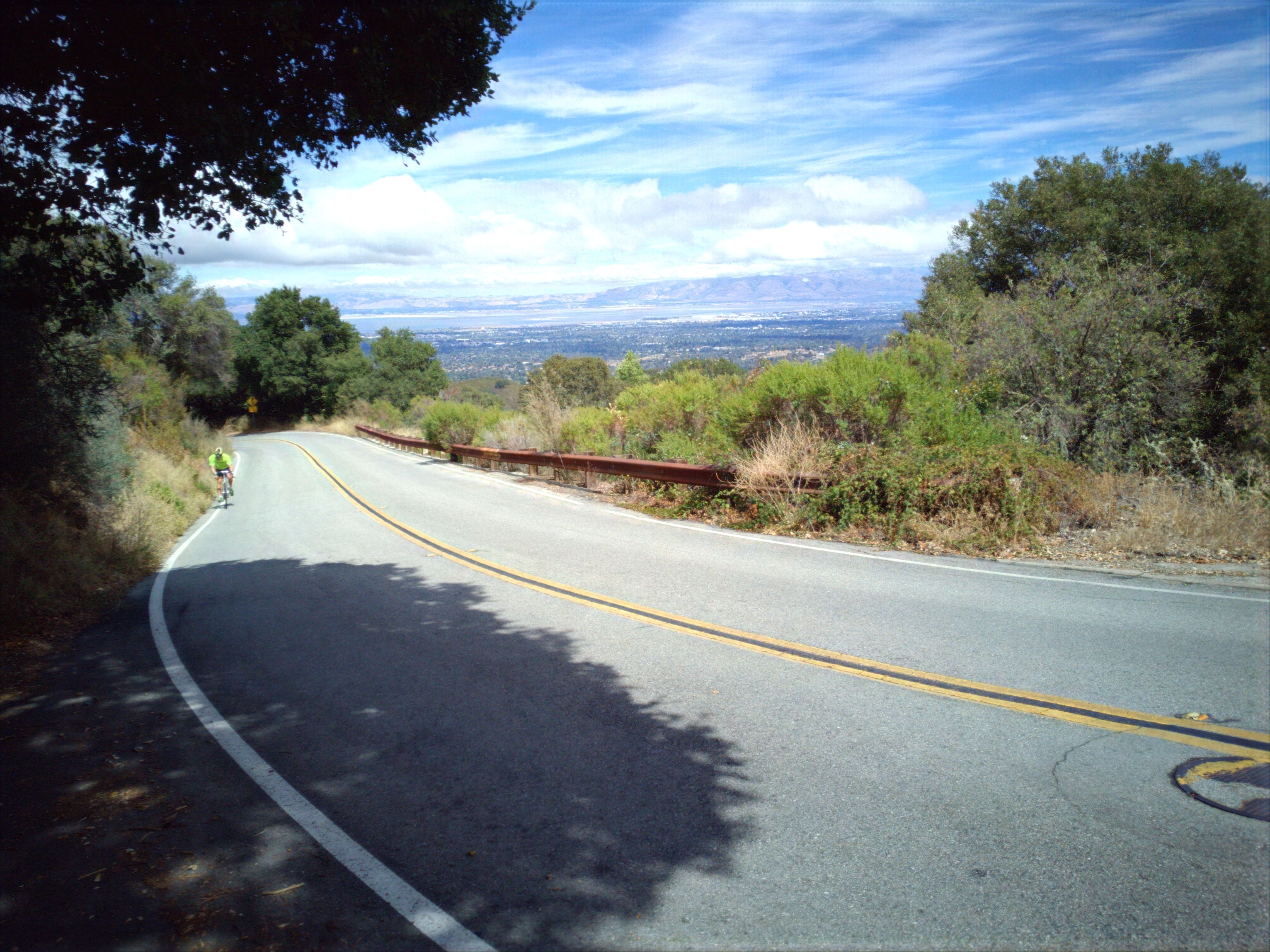} };
        \zoombox[magnification=10, color code=red]{0.175,0.8}
        \zoombox[magnification=10, color code=yellow]{0.51,0.68}
        \zoombox[magnification=10, color code=orange]{0.175,0.51}
        \zoombox[magnification=10,color code=lime]{0.82,0.71}
    \end{tikzpicture}\vspace{-5px}\caption{DSLR}\vspace{3px} \end{subfigure} \begin{subfigure}[b]{0.11\textwidth}
    \begin{tikzpicture}[zoomboxarray, zoomboxes below]
        \node [image node] { \includegraphics[width=0.95\textwidth]{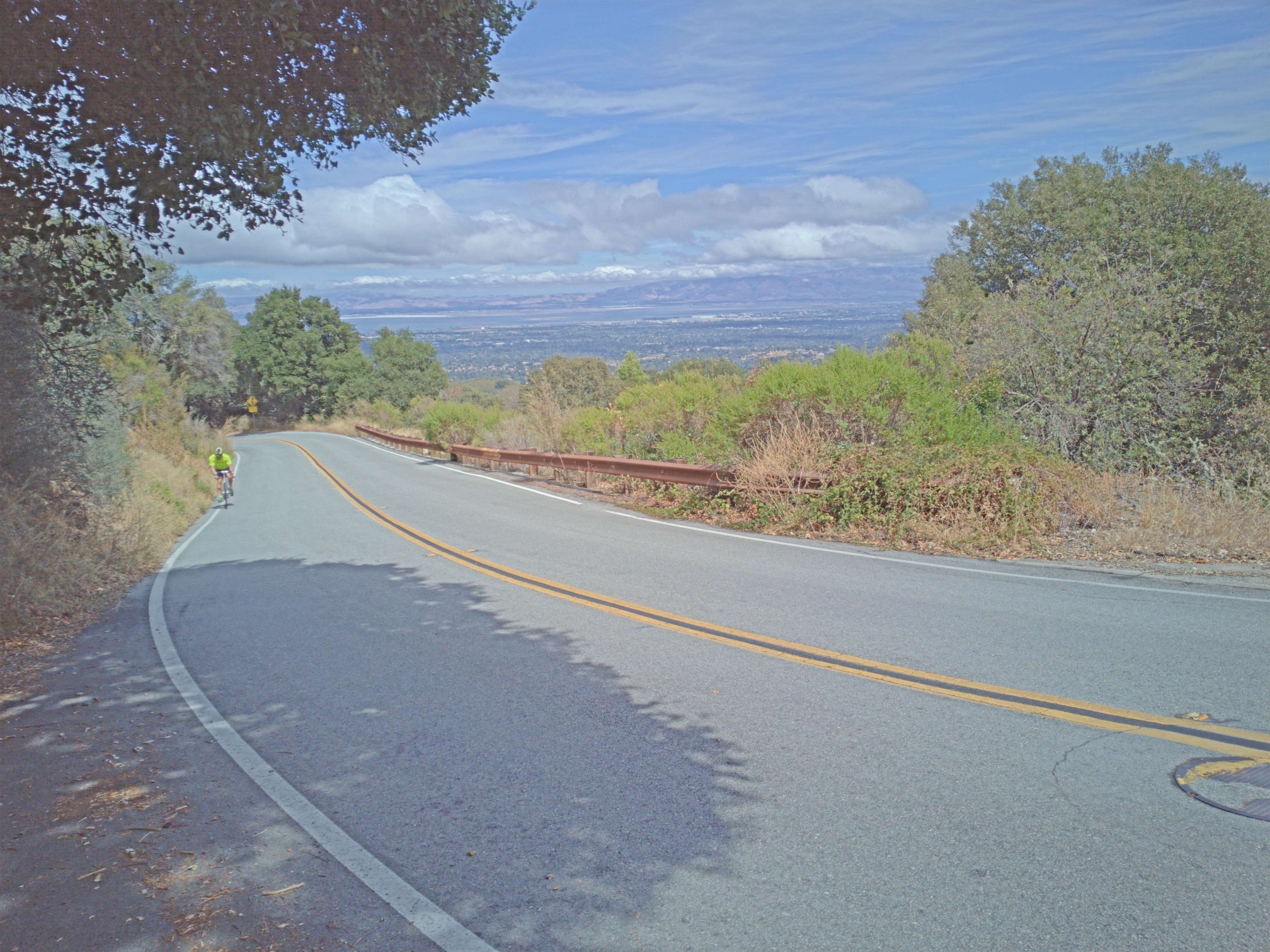} };
        \zoombox[magnification=10, color code=red]{0.175,0.8}
        \zoombox[magnification=10, color code=yellow]{0.51,0.68}
        \zoombox[magnification=10, color code=orange]{0.175,0.51}
        \zoombox[magnification=10,color code=lime]{0.82,0.71}
    \end{tikzpicture}\vspace{-5px}\caption{TBEFN}\vspace{3px} \end{subfigure} \begin{subfigure}[b]{0.11\textwidth}
    \begin{tikzpicture}[zoomboxarray, zoomboxes below]
        \node [image node] { \includegraphics[width=0.95\textwidth]{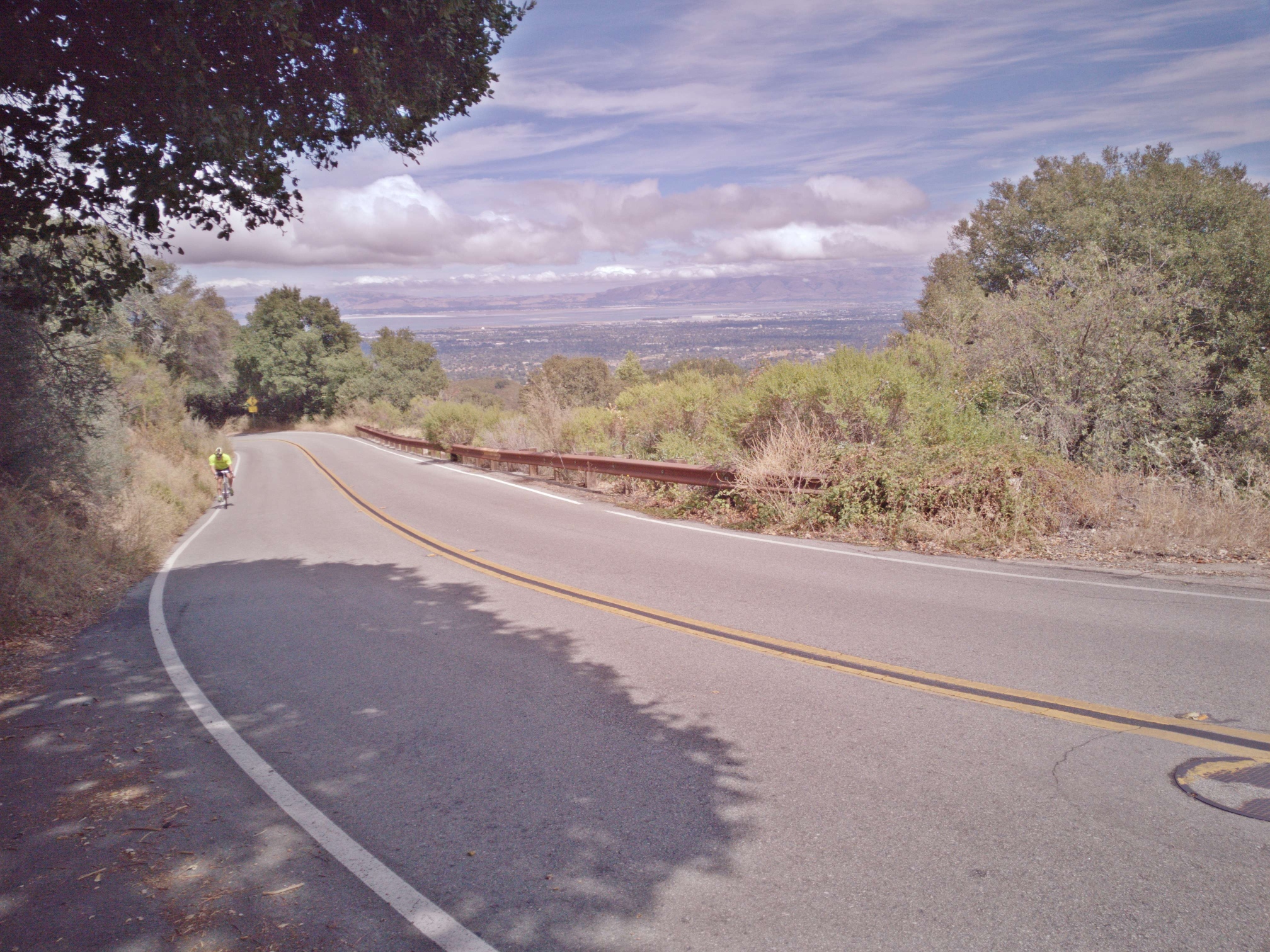} };
        \zoombox[magnification=10, color code=red]{0.175,0.8}
        \zoombox[magnification=10, color code=yellow]{0.51,0.68}
        \zoombox[magnification=10, color code=orange]{0.175,0.51}
        \zoombox[magnification=10,color code=lime]{0.82,0.71}
    \end{tikzpicture}\vspace{-5px}\caption{Zero-DCE++}\vspace{3px} \end{subfigure} \begin{subfigure}[b]{0.11\textwidth}
    \begin{tikzpicture}[zoomboxarray, zoomboxes below]
        \node [image node] { \includegraphics[width=0.95\textwidth]{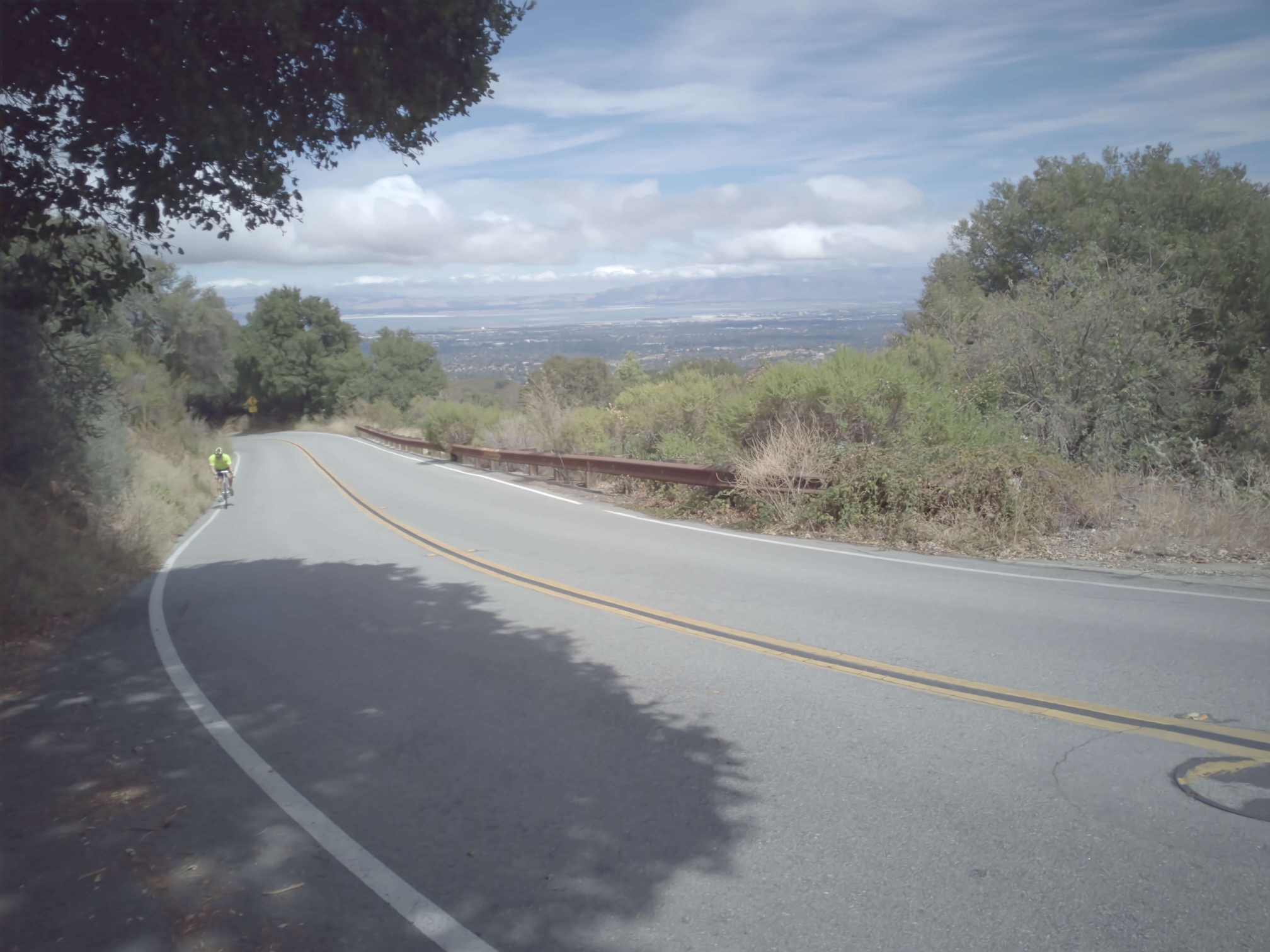} };
        \zoombox[magnification=10, color code=red]{0.175,0.8}
        \zoombox[magnification=10, color code=yellow]{0.51,0.68}
        \zoombox[magnification=10, color code=orange]{0.175,0.51}
        \zoombox[magnification=10,color code=lime]{0.82,0.71}
    \end{tikzpicture}\vspace{-5px}\caption{FFDNet+GC}\vspace{3px} \end{subfigure} \begin{subfigure}[b]{0.11\textwidth}
    \begin{tikzpicture}[zoomboxarray, zoomboxes below]
        \node [image node] { \includegraphics[width=0.95\textwidth]{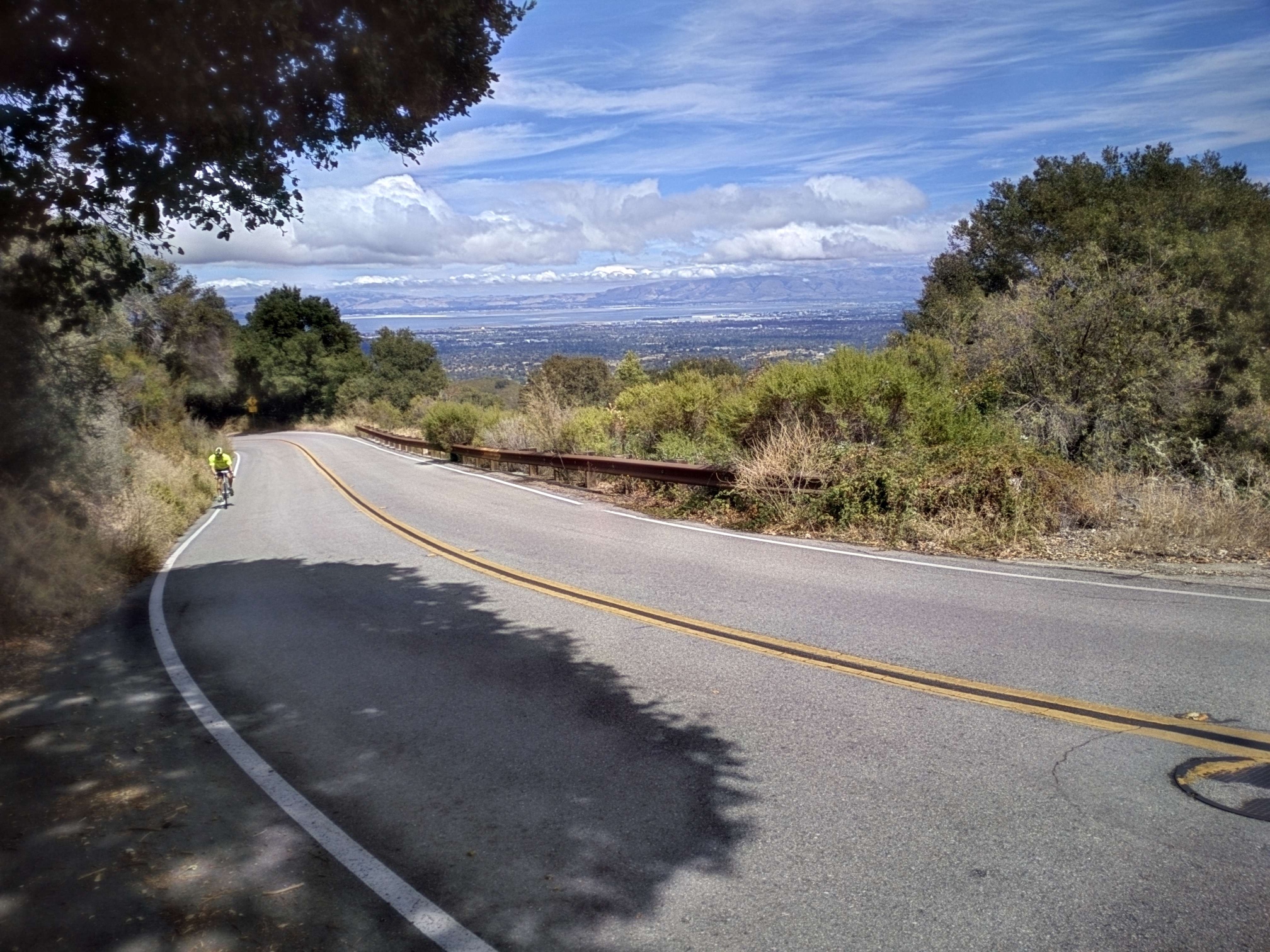} };
        \zoombox[magnification=10, color code=red]{0.175,0.8}
        \zoombox[magnification=10, color code=yellow]{0.51,0.68}
        \zoombox[magnification=10, color code=orange]{0.175,0.51}
        \zoombox[magnification=10,color code=lime]{0.82,0.71}
    \end{tikzpicture}\vspace{-5px}\caption{DFTL (Ours)}\vspace{3px} \end{subfigure} \begin{subfigure}[b]{0.11\textwidth}
    \begin{tikzpicture}[zoomboxarray, zoomboxes below]
        \node [image node] { \includegraphics[width=0.95\textwidth]{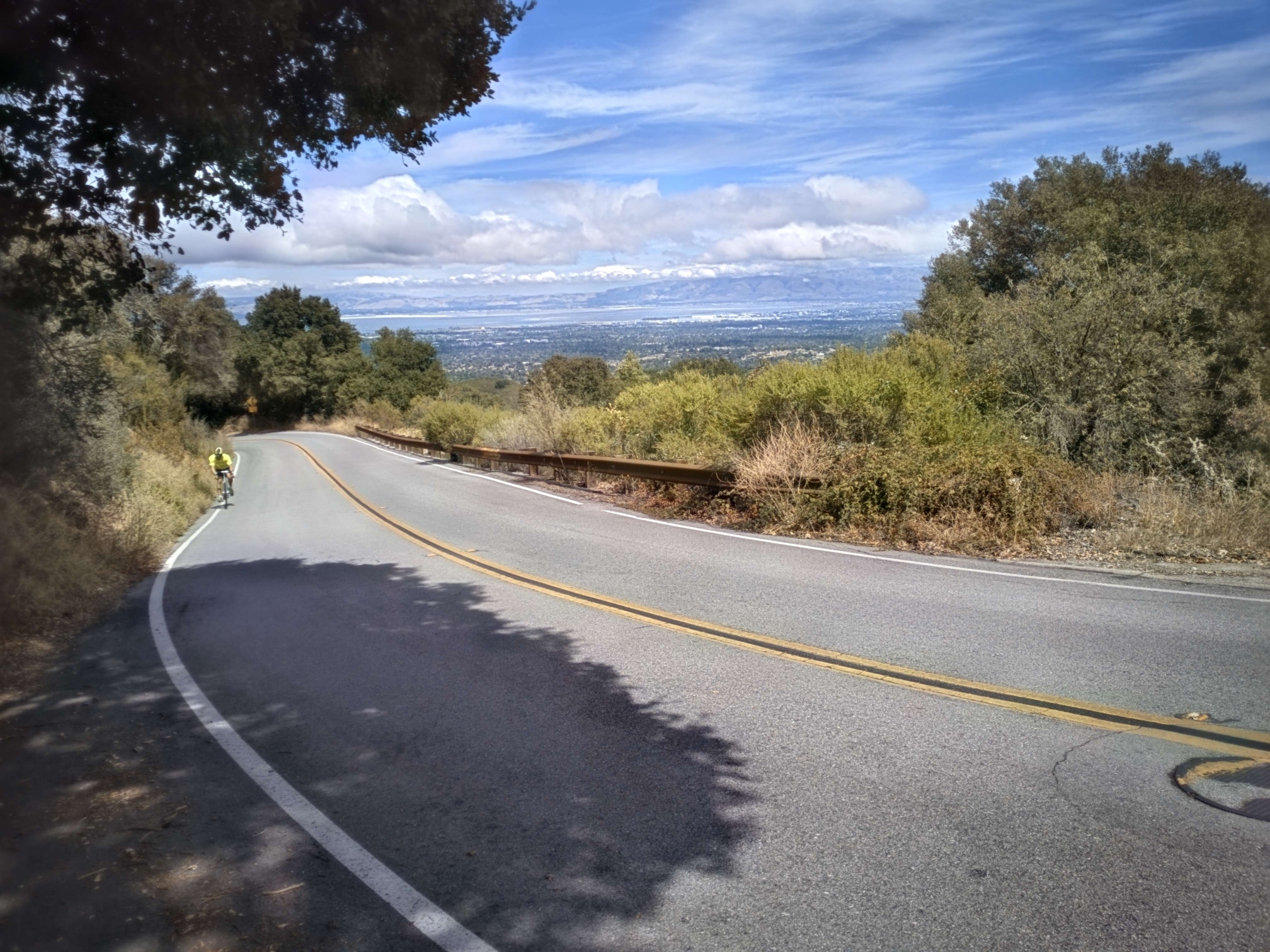} };
        \zoombox[magnification=10, color code=red]{0.175,0.8}
        \zoombox[magnification=10, color code=yellow]{0.51,0.68}
        \zoombox[magnification=10, color code=orange]{0.175,0.51}
        \zoombox[magnification=10,color code=lime]{0.82,0.71}
    \end{tikzpicture}\vspace{-5px}\caption{TFDL (Ours)}\vspace{3px} \end{subfigure} \begin{subfigure}[b]{0.11\textwidth}
    \begin{tikzpicture}[zoomboxarray, zoomboxes below]
        \node [image node] { \includegraphics[width=0.95\textwidth]{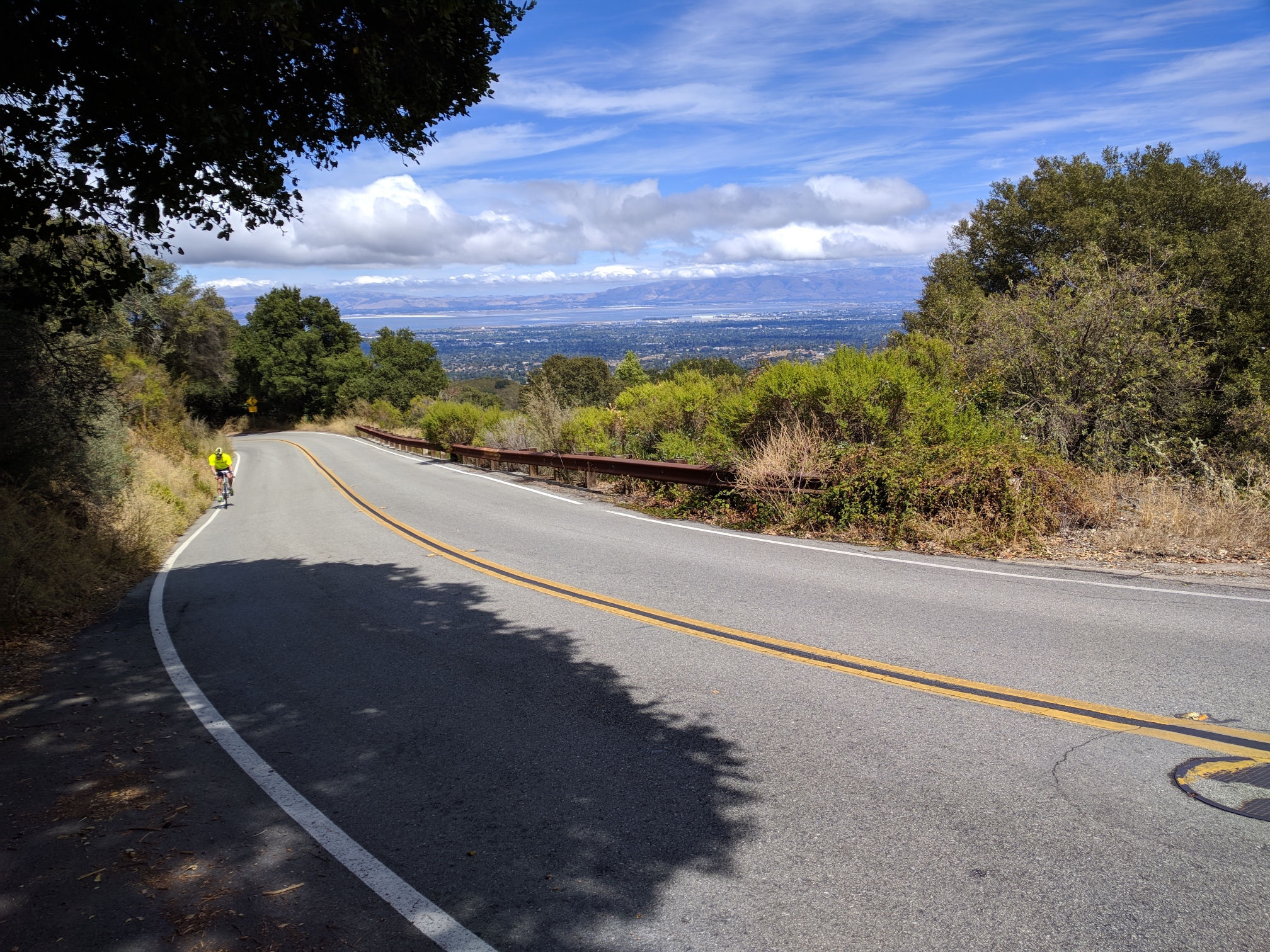} };
        \zoombox[magnification=10, color code=red]{0.175,0.8}
        \zoombox[magnification=10, color code=yellow]{0.51,0.68}
        \zoombox[magnification=10, color code=orange]{0.175,0.51}
        \zoombox[magnification=10,color code=lime]{0.82,0.71}
    \end{tikzpicture}\vspace{-5px}\caption{Ground-truth}\vspace{3px} \end{subfigure}
  \caption{Qualitative comparisons between methods on examples from the HDR+ testing set. Four sub-regions selected from each example are zoomed in 10 times to show more details. Images here have been down-sampled for better viewing experience.}
\label{fig:example}
\end{figure*}

\section{Experiments}

In this section, we will demonstrate the effectiveness of our proposed framework through both qualitative and quantitative evaluation on several HDR image datasets.

\subsection{Training Settings and Dataset}

Our experiments are run on a single Nvidia Tesla GPU. Detailed training settings for the three training phases for our joint framework are summarized in Table \ref{tab:setting}. Adam optimizers with $\beta_1 =0.9$ and $\beta_2=0.999$ are adopted for all 3 training phases. We empirically choose the hyper-parameters in the loss functions (Equation \ref{eq:loss1-1} and Equation \ref{eq:loss2-1}) to be $\lambda_0=2, \lambda_1=2, \lambda_2=2, \lambda_3=1, \lambda_d=1$.

\begin{table}[ht!]
\begin{center}
\begin{tabular}{|c|c|c|c|}
\hline
Phase & Batch Size & Learning Rate  & Epochs \\
\hline\hline
1 & 16 & 1e-5 & 500\\
\hline
2 & 16 & 1e-5 & 500\\
\hline
3 & 16 & 1e-6 & 1000\\
\hline
\end{tabular}
\end{center}
\caption{Training settings for the three phases.}
\label{tab:setting}
\end{table}

\begin{figure*}
\captionsetup[subfigure]{labelformat=empty}
  \centering
    \begin{subfigure}[b]{0.11\textwidth}
    \begin{tikzpicture}[zoomboxarray, zoomboxes below]
        \node [image node] { \includegraphics[width=0.95\textwidth]{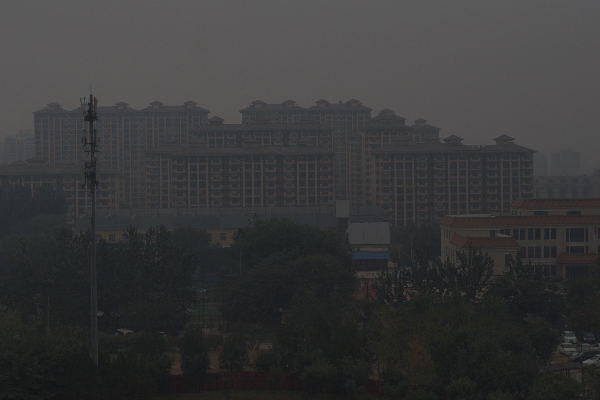} };
        \zoombox[magnification=5, color code=red]{0.175,0.7}
        \zoombox[magnification=5, color code=yellow]{0.61,0.6}
        \zoombox[magnification=5, color code=orange]{0.25,0.29}
        \zoombox[magnification=5,color code=lime]{0.82,0.65}
    \end{tikzpicture}\vspace{-5px}\caption{Input}\vspace{3px} \end{subfigure} \begin{subfigure}[b]{0.11\textwidth}
    \begin{tikzpicture}[zoomboxarray, zoomboxes below]
        \node [image node] { \includegraphics[width=0.95\textwidth]{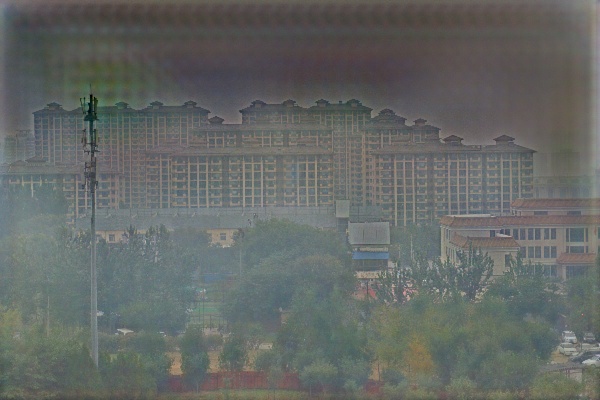} };
        \zoombox[magnification=5, color code=red]{0.175,0.7}
        \zoombox[magnification=5, color code=yellow]{0.61,0.6}
        \zoombox[magnification=5, color code=orange]{0.25,0.29}
        \zoombox[magnification=5,color code=lime]{0.82,0.65}
    \end{tikzpicture}\vspace{-5px}\caption{DSLR}\vspace{3px} \end{subfigure} \begin{subfigure}[b]{0.11\textwidth}
    \begin{tikzpicture}[zoomboxarray, zoomboxes below]
        \node [image node] { \includegraphics[width=0.95\textwidth]{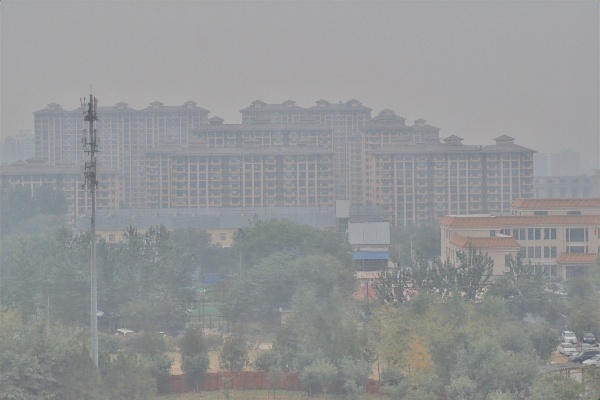} };
        \zoombox[magnification=5, color code=red]{0.175,0.7}
        \zoombox[magnification=5, color code=yellow]{0.61,0.6}
        \zoombox[magnification=5, color code=orange]{0.25,0.29}
        \zoombox[magnification=5,color code=lime]{0.82,0.65}
    \end{tikzpicture}\vspace{-5px}\caption{TBEFN}\vspace{3px} \end{subfigure} \begin{subfigure}[b]{0.11\textwidth}
    \begin{tikzpicture}[zoomboxarray, zoomboxes below]
        \node [image node] { \includegraphics[width=0.95\textwidth]{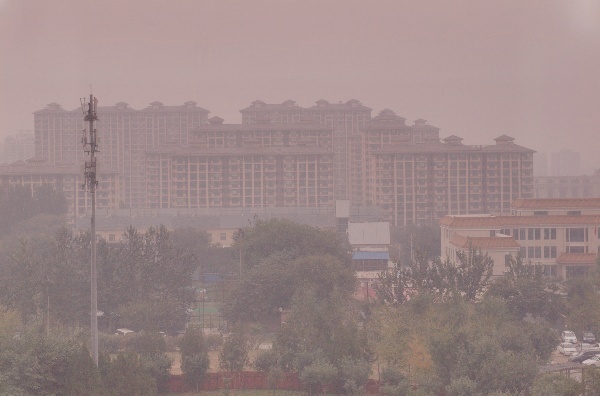} };
        \zoombox[magnification=5, color code=red]{0.175,0.7}
        \zoombox[magnification=5, color code=yellow]{0.61,0.6}
        \zoombox[magnification=5, color code=orange]{0.25,0.29}
        \zoombox[magnification=5,color code=lime]{0.82,0.65}
    \end{tikzpicture}\vspace{-5px}\caption{Zero-DCE++}\vspace{3px} \end{subfigure} \begin{subfigure}[b]{0.11\textwidth}
    \begin{tikzpicture}[zoomboxarray, zoomboxes below]
        \node [image node] { \includegraphics[width=0.95\textwidth]{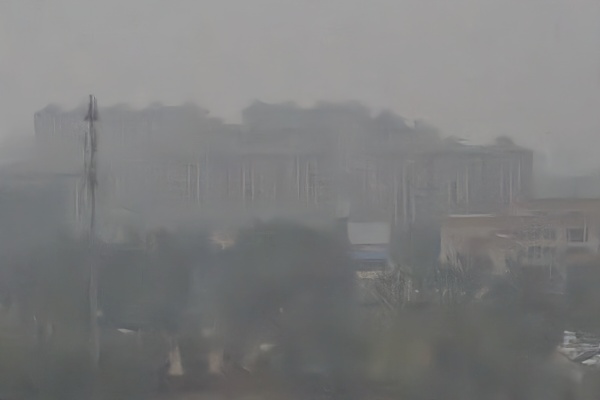} };
        \zoombox[magnification=5, color code=red]{0.175,0.7}
        \zoombox[magnification=5, color code=yellow]{0.61,0.6}
        \zoombox[magnification=5, color code=orange]{0.25,0.29}
        \zoombox[magnification=5,color code=lime]{0.82,0.65}
    \end{tikzpicture}\vspace{-5px}\caption{FFDNet+GC}\vspace{3px} \end{subfigure} \begin{subfigure}[b]{0.11\textwidth}
    \begin{tikzpicture}[zoomboxarray, zoomboxes below]
        \node [image node] { \includegraphics[width=0.95\textwidth]{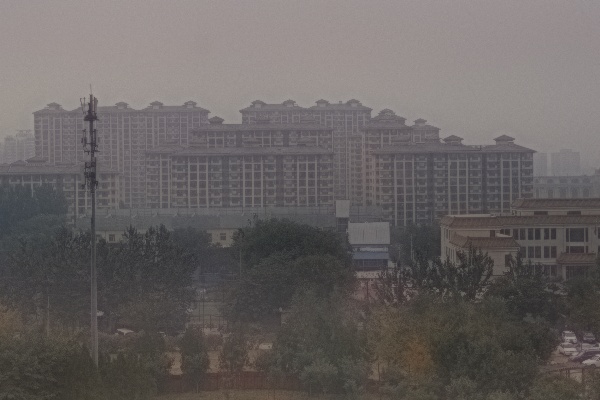} };
        \zoombox[magnification=5, color code=red]{0.175,0.7}
        \zoombox[magnification=5, color code=yellow]{0.61,0.6}
        \zoombox[magnification=5, color code=orange]{0.25,0.29}
        \zoombox[magnification=5,color code=lime]{0.82,0.65}
    \end{tikzpicture}\vspace{-5px}\caption{DFTL (Ours)}\vspace{3px} \end{subfigure} \begin{subfigure}[b]{0.11\textwidth}
    \begin{tikzpicture}[zoomboxarray, zoomboxes below]
        \node [image node] { \includegraphics[width=0.95\textwidth]{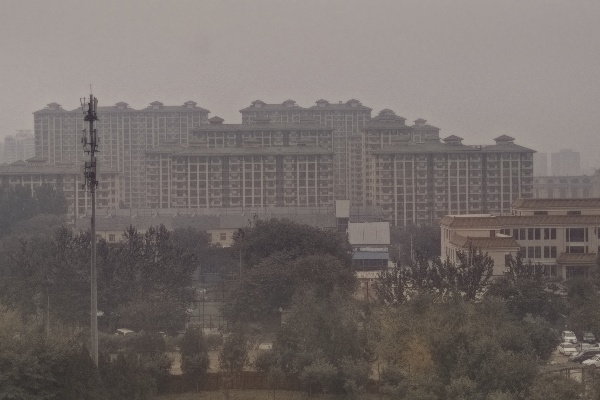} };
        \zoombox[magnification=5, color code=red]{0.175,0.7}
        \zoombox[magnification=5, color code=yellow]{0.61,0.6}
        \zoombox[magnification=5, color code=orange]{0.25,0.29}
        \zoombox[magnification=5,color code=lime]{0.82,0.65}
    \end{tikzpicture}\vspace{-5px}\caption{TFDL (Ours)}\vspace{3px} \end{subfigure} \begin{subfigure}[b]{0.11\textwidth}
    \begin{tikzpicture}[zoomboxarray, zoomboxes below]
        \node [image node] { \includegraphics[width=0.95\textwidth]{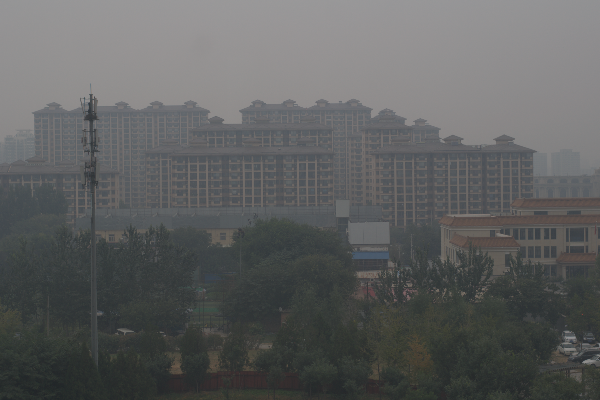} };
        \zoombox[magnification=5, color code=red]{0.175,0.7}
        \zoombox[magnification=5, color code=yellow]{0.61,0.6}
        \zoombox[magnification=5, color code=orange]{0.25,0.29}
        \zoombox[magnification=5,color code=lime]{0.82,0.65}
    \end{tikzpicture}\vspace{-5px}\caption{GT}\vspace{3px} \end{subfigure}
    \begin{subfigure}[b]{0.11\textwidth}
    \begin{tikzpicture}[zoomboxarray, zoomboxes below]
        \node [image node] { \includegraphics[width=0.95\textwidth]{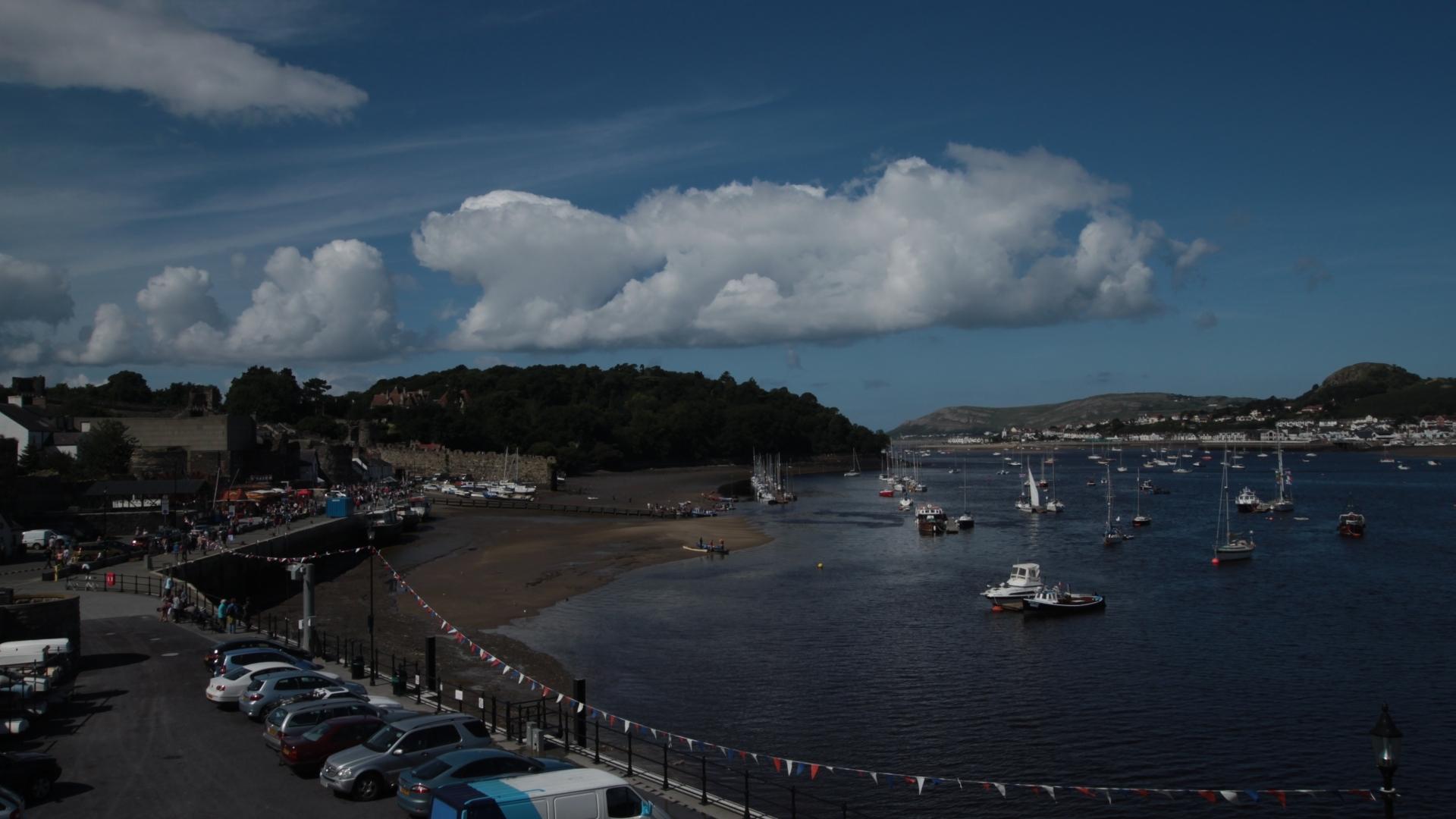} };
        \zoombox[magnification=5, color code=red]{0.2,0.4}
        \zoombox[magnification=5, color code=yellow]{0.7,0.3}
        \zoombox[magnification=5, color code=orange]{0.23,0.17}
        \zoombox[magnification=5,color code=lime]{0.72,0.7}
    \end{tikzpicture}\vspace{-5px}\caption{Input}\vspace{3px} \end{subfigure} \begin{subfigure}[b]{0.11\textwidth}
    \begin{tikzpicture}[zoomboxarray, zoomboxes below]
        \node [image node] { \includegraphics[width=0.95\textwidth]{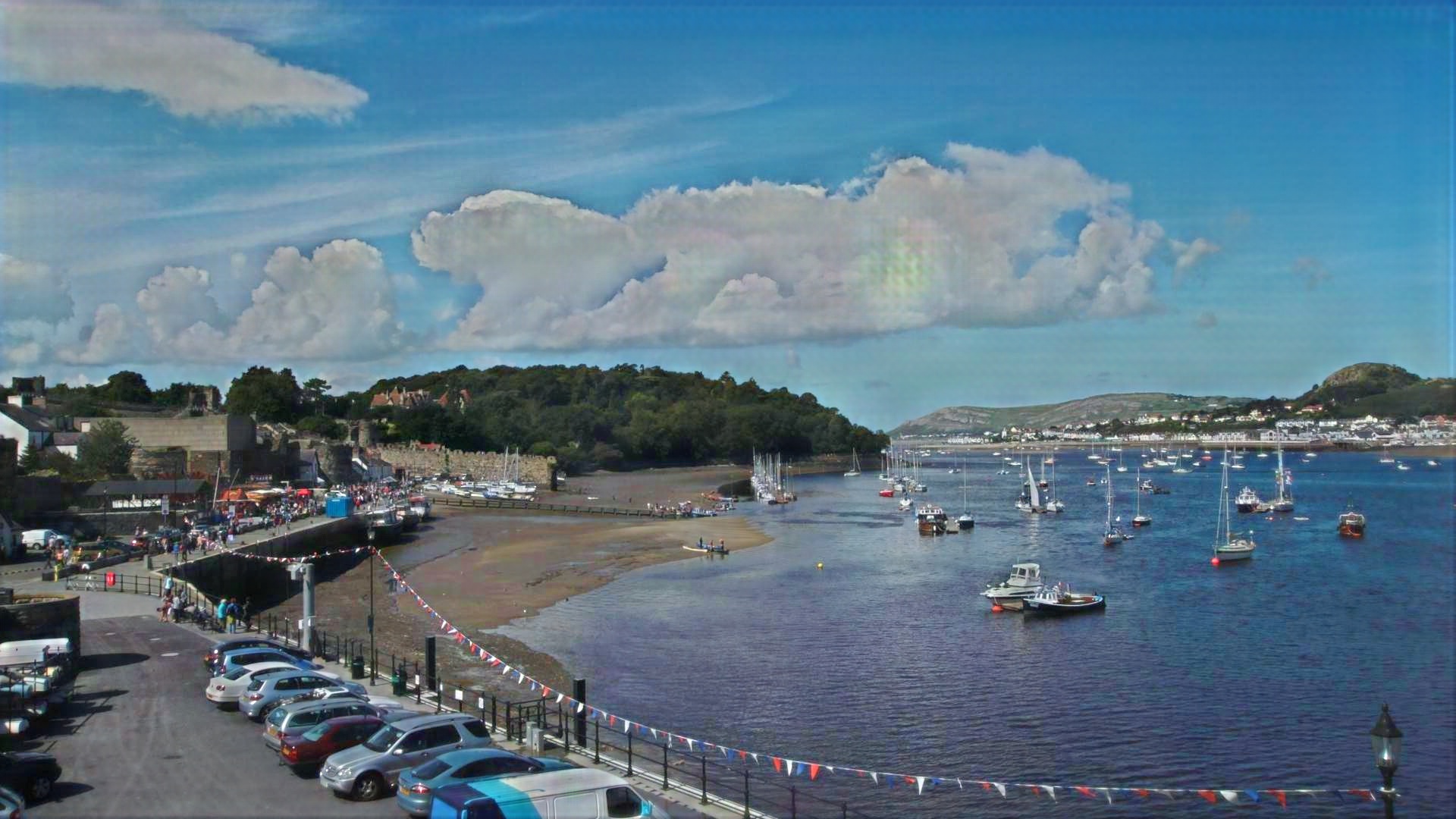} };
        \zoombox[magnification=5, color code=red]{0.2,0.4}
        \zoombox[magnification=5, color code=yellow]{0.7,0.3}
        \zoombox[magnification=5, color code=orange]{0.23,0.17}
        \zoombox[magnification=5,color code=lime]{0.72,0.7}
    \end{tikzpicture}\vspace{-5px}\caption{DSLR}\vspace{3px} \end{subfigure} \begin{subfigure}[b]{0.11\textwidth}
    \begin{tikzpicture}[zoomboxarray, zoomboxes below]
        \node [image node] { \includegraphics[width=0.95\textwidth]{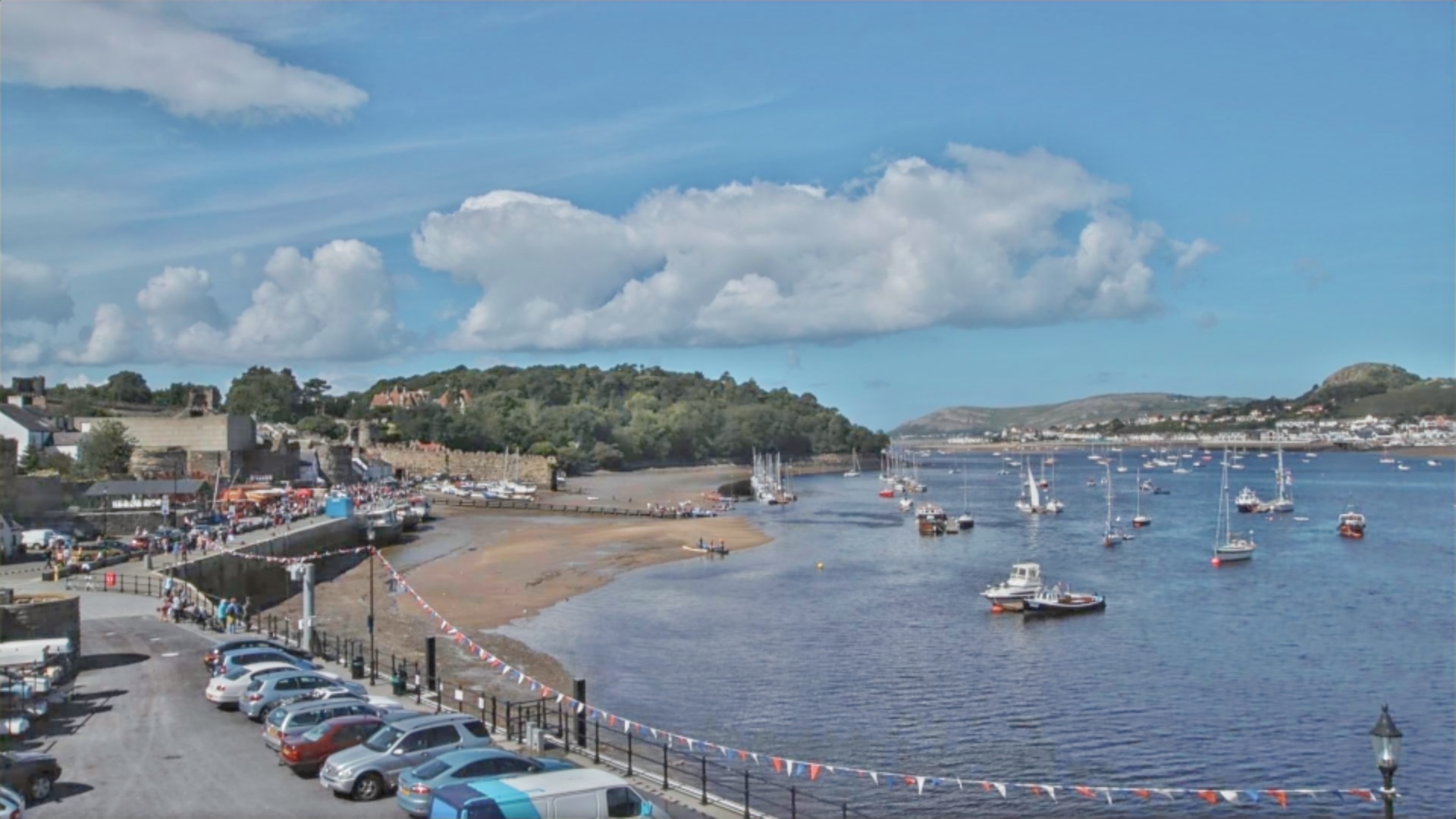} };
        \zoombox[magnification=5, color code=red]{0.2,0.4}
        \zoombox[magnification=5, color code=yellow]{0.7,0.3}
        \zoombox[magnification=5, color code=orange]{0.23,0.17}
        \zoombox[magnification=5,color code=lime]{0.72,0.7}
    \end{tikzpicture}\vspace{-5px}\caption{TBEFN}\vspace{3px} \end{subfigure} \begin{subfigure}[b]{0.11\textwidth}
    \begin{tikzpicture}[zoomboxarray, zoomboxes below]
        \node [image node] { \includegraphics[width=0.95\textwidth]{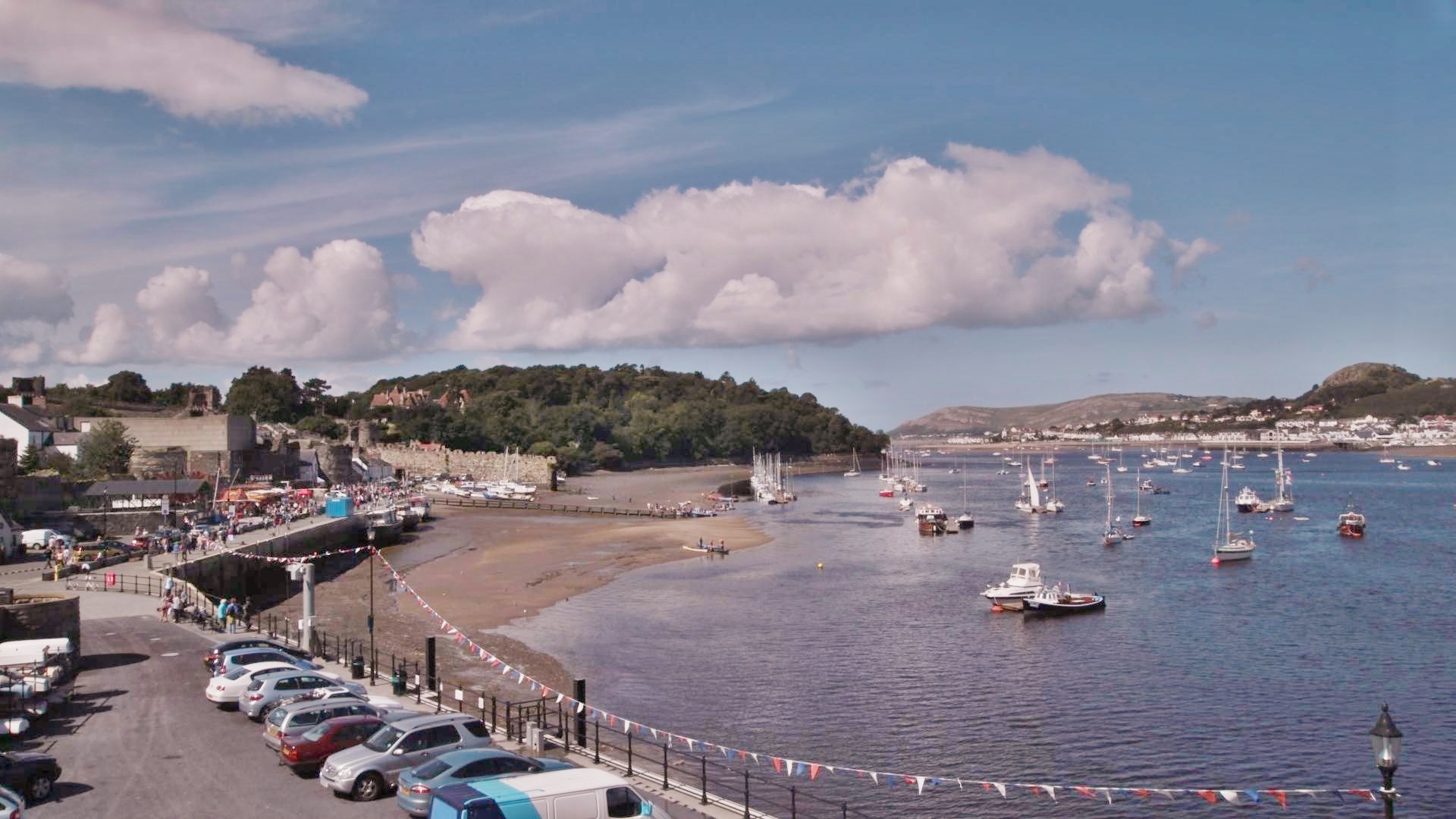} };
        \zoombox[magnification=5, color code=red]{0.2,0.4}
        \zoombox[magnification=5, color code=yellow]{0.7,0.3}
        \zoombox[magnification=5, color code=orange]{0.23,0.17}
        \zoombox[magnification=5,color code=lime]{0.72,0.7}
    \end{tikzpicture}\vspace{-5px}\caption{Zero-DCE++}\vspace{3px} \end{subfigure} \begin{subfigure}[b]{0.11\textwidth}
    \begin{tikzpicture}[zoomboxarray, zoomboxes below]
        \node [image node] { \includegraphics[width=0.95\textwidth]{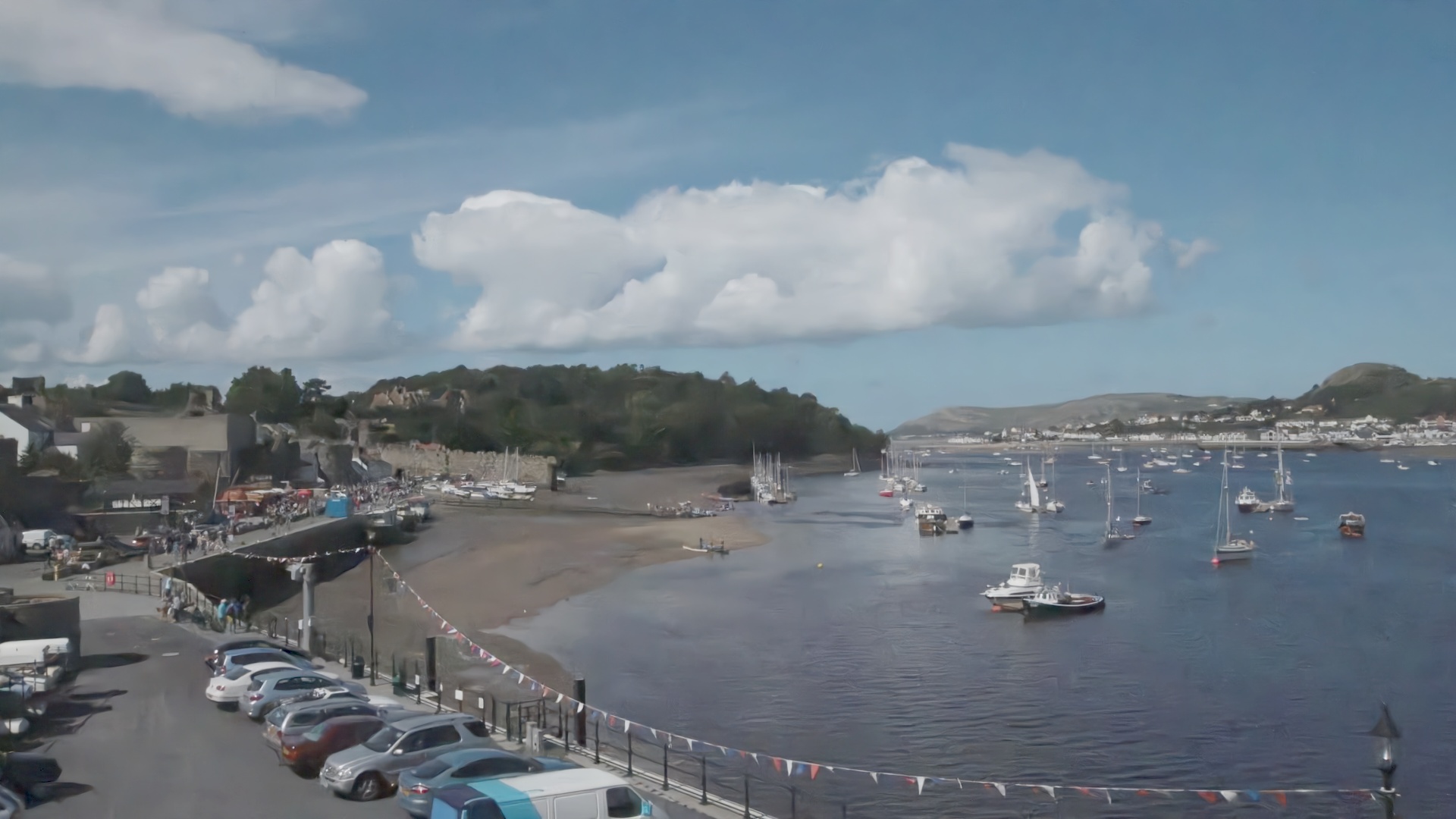} };
        \zoombox[magnification=5, color code=red]{0.2,0.4}
        \zoombox[magnification=5, color code=yellow]{0.7,0.3}
        \zoombox[magnification=5, color code=orange]{0.23,0.17}
        \zoombox[magnification=5,color code=lime]{0.72,0.7}
    \end{tikzpicture}\vspace{-5px}\caption{FFDNet+GC}\vspace{3px} \end{subfigure} \begin{subfigure}[b]{0.11\textwidth}
    \begin{tikzpicture}[zoomboxarray, zoomboxes below]
        \node [image node] { \includegraphics[width=0.95\textwidth]{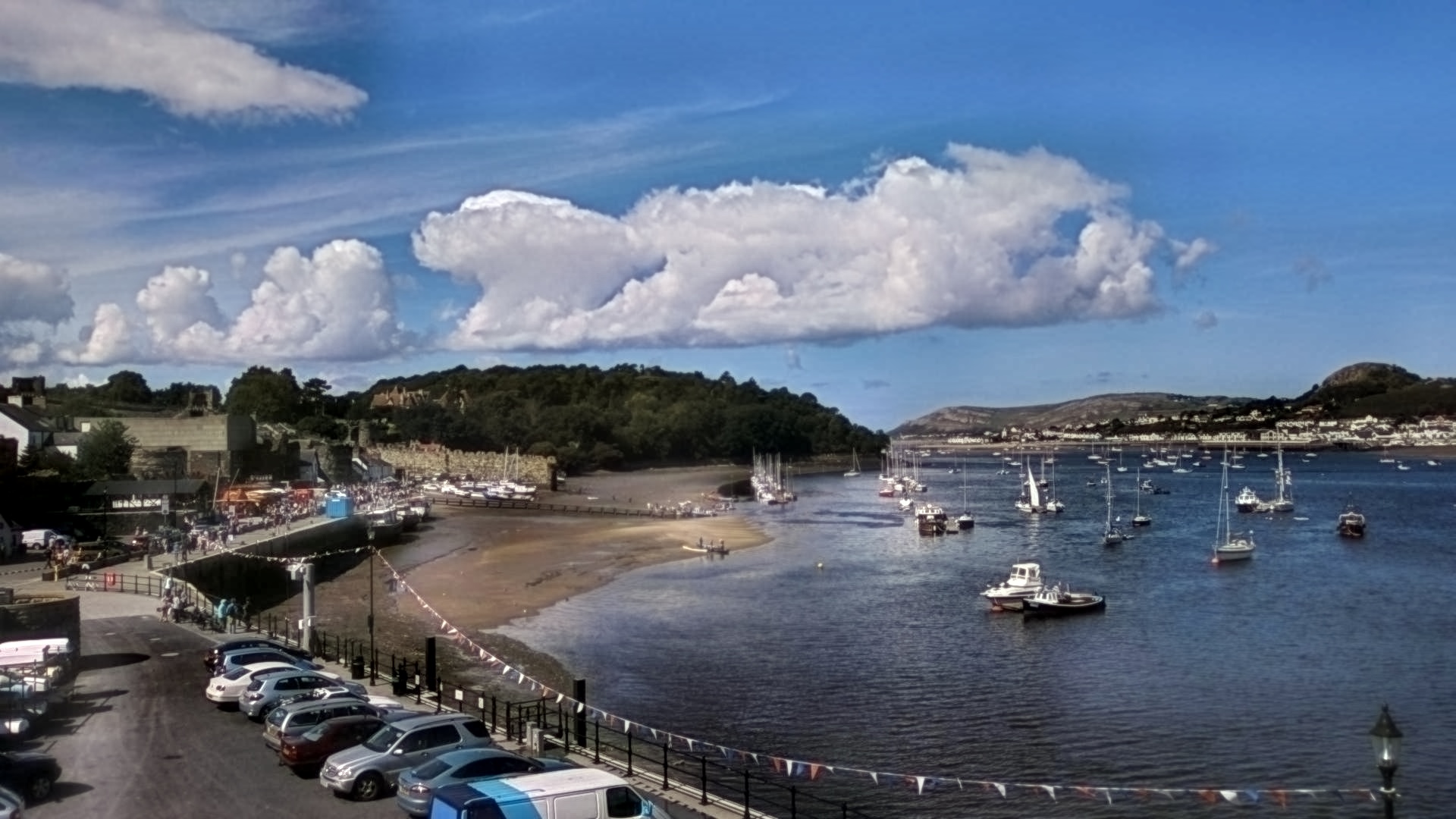} };
        \zoombox[magnification=5, color code=red]{0.2,0.4}
        \zoombox[magnification=5, color code=yellow]{0.7,0.3}
        \zoombox[magnification=5, color code=orange]{0.23,0.17}
        \zoombox[magnification=5,color code=lime]{0.72,0.7}
    \end{tikzpicture}\vspace{-5px}\caption{DFTL (Ours)}\vspace{3px} \end{subfigure} \begin{subfigure}[b]{0.11\textwidth}
    \begin{tikzpicture}[zoomboxarray, zoomboxes below]
        \node [image node] { \includegraphics[width=0.95\textwidth]{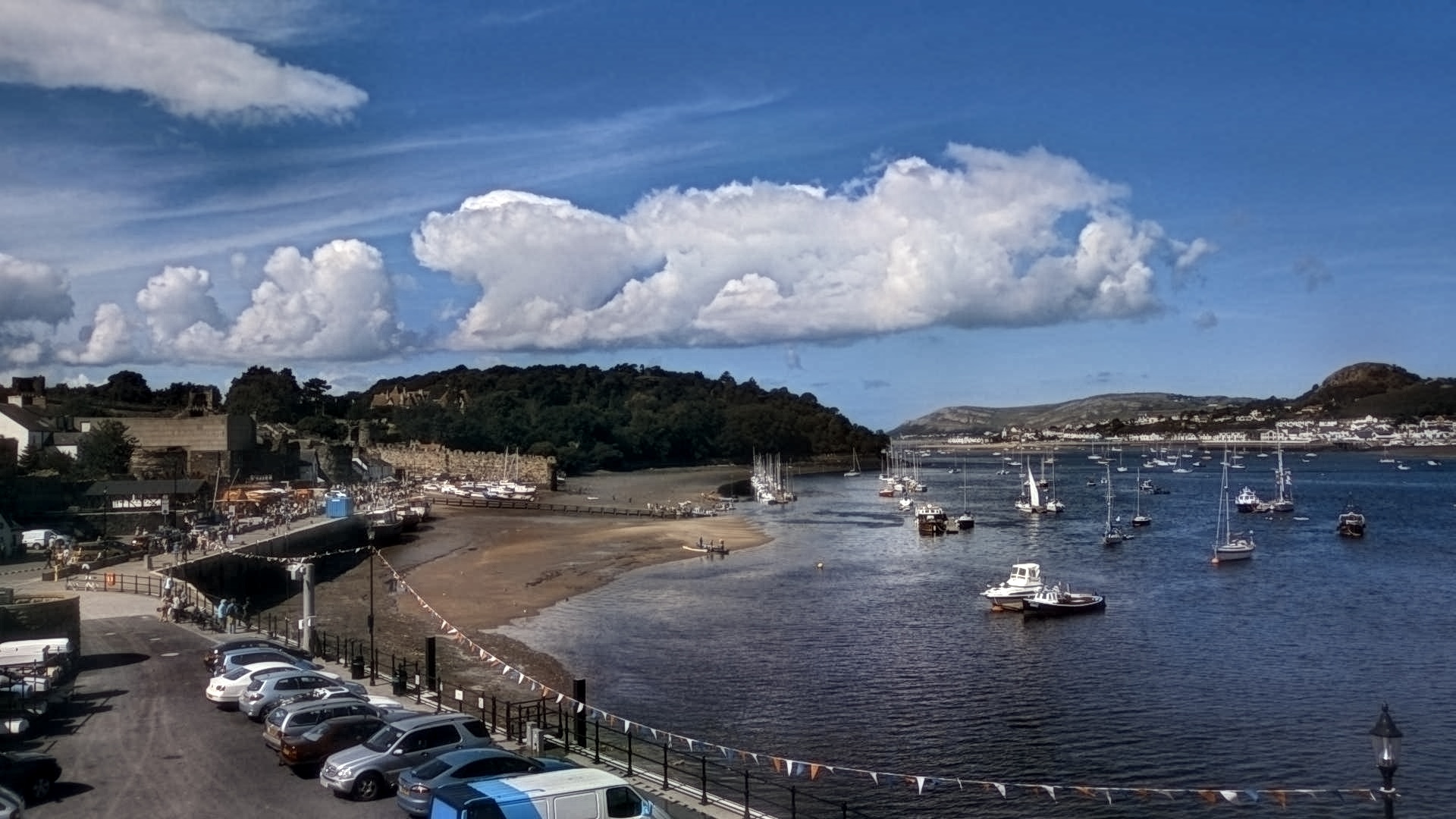} };
        \zoombox[magnification=5, color code=red]{0.2,0.4}
        \zoombox[magnification=5, color code=yellow]{0.7,0.3}
        \zoombox[magnification=5, color code=orange]{0.23,0.17}
        \zoombox[magnification=5,color code=lime]{0.72,0.7}
    \end{tikzpicture}\vspace{-5px}\caption{TFDL (Ours)}\vspace{3px} \end{subfigure}
     \begin{subfigure}[b]{0.11\textwidth}
    \begin{tikzpicture}[zoomboxarray, zoomboxes below]
        \node [image node] { \includegraphics[width=0.95\textwidth]{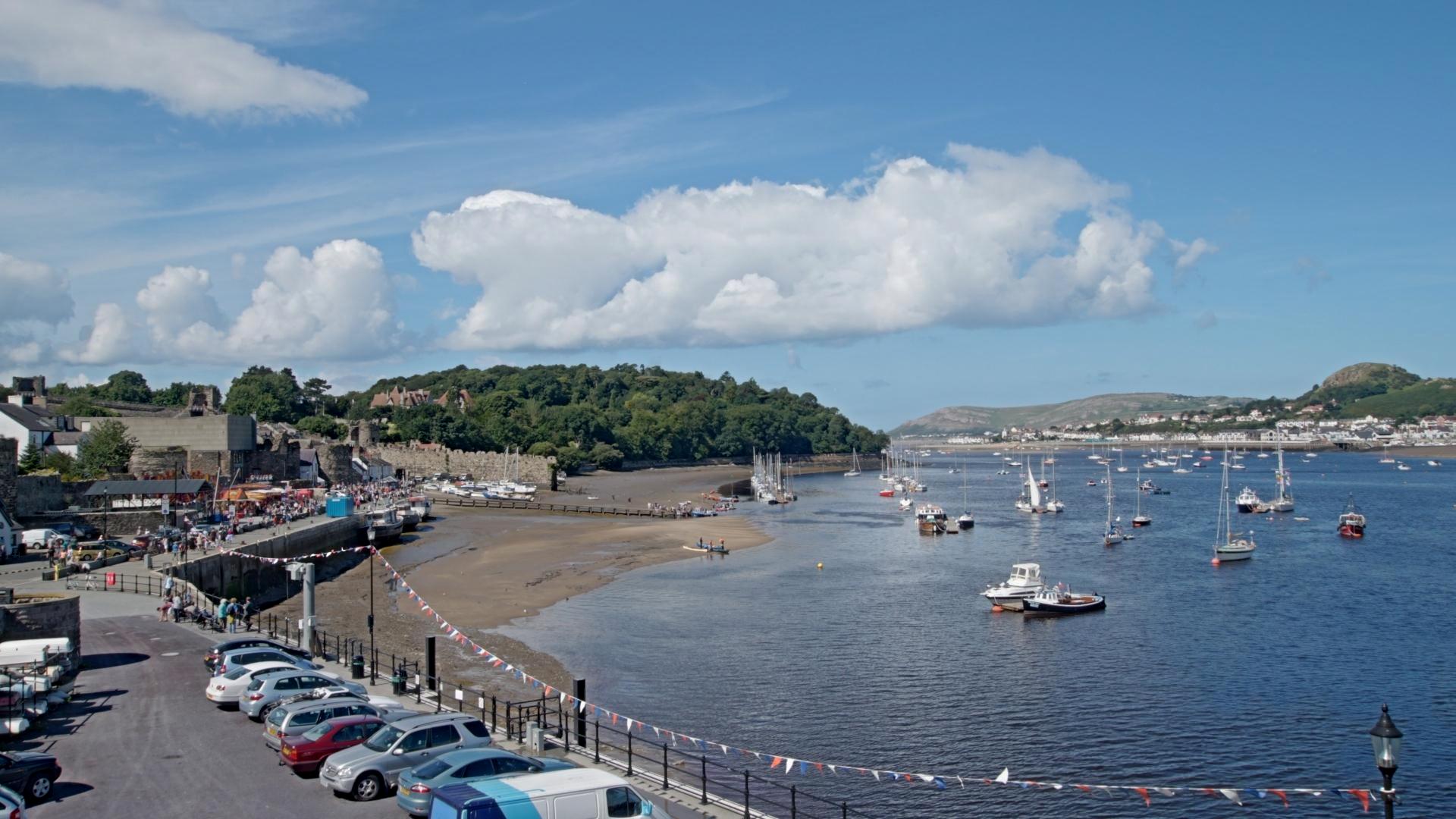} };
        \zoombox[magnification=5, color code=red]{0.2,0.4}
        \zoombox[magnification=5, color code=yellow]{0.7,0.3}
        \zoombox[magnification=5, color code=orange]{0.23,0.17}
        \zoombox[magnification=5,color code=lime]{0.72,0.7}
    \end{tikzpicture}\vspace{-5px}\caption{GT}\vspace{3px} \end{subfigure}
  \caption{Qualitative comparisons between methods on examples from the LOL dataset (the first row) and SCIE dataset (the second row). Four sub-regions selected from each example are zoomed in 5 times to show more details.}
\label{fig:example1}
\end{figure*}

We use the Google HDR+ dataset \cite{hasinoff2016burst} as our training dataset, as it provides both the merged intermediate ground-truth image ( denoised using burst captures) and the final ground-truth image (denoised and tone-mapped) for calculating our loss functions. From the HDR+ dataset, we randomly select 2000 scenes for training, 200 scenes for validation, and 1300 scenes for testing and benchmarking. 

The input images we used from the HDR+ dataset are the reference frames from the $N$ input burst photos of each individual scene, and the merged intermediate ground-truth images are the intermediate raw results of aligning and merging their corresponding input burst. For simplicity, we pre-processed the raw input images and raw merged intermediate images by demosaicing, color correction, and gamma correction to convert them to sRGB color space before using them. Note that the ground-truth images in the HDR+ dataset are obtained through a series of processing steps in the finishing pipelines besides denoising and tone-mapping. However, including all the finishing steps in our framework would overly complicate our method, so the rest of the processing steps in the finishing pipeline are not explicitly performed in our framework when obtaining the final tone-mapped and denoised output, and we can assume that these operations are implicitly learned through the tone-mapping modules and denoising modules in our joint multi-scale framework.

Since the provided final ground-truth images are cropped, we need to align the input images and the merged intermediate ground-truth images with their final counterparts. For convenience, we used the Scale-Invariant Feature Transform (SIFT) \cite{Burger2016} to align and crop the input and intermediate images as a prepossessing step.

Our joint framework takes input with a size of $224\times 224$ pixels, so we cannot use the full images to train our model. Instead, we split all 2000 input images into tiles of size $224\times 224$ pixels, along with their intermediate merged ground-truth image and final ground-truth image.

\subsection{Results}

\subsubsection{Full-Reference Comparison}

To quantitatively evaluate the performance of our proposed joint tone-mapping and denoising framework, we measured the SSIM, PSNR, LPIPS \cite{zhang2018perceptual}, and Tone-Mapping Quality Index (TMQI) \cite{tmqi} on the HDR+ testing set (1300 scenes), as well as the LOL training dataset (485 scenes) \cite{Wei2018DeepRD} and part 1 of the SCIE dataset (360 scenes) \cite{Cai2018deep}. Note that our models are neither trained nor fine-tuned on any subsets of the LOL dataset or the SCIE dataset. For the HDR+ dataset, we used the reference frame as the input and for the SCIE dataset, we used the first image (the image with the lowest exposure) of every scene as the input. We tested several state-of-the-art image enhancement methods including DSLR\cite{DSLR}, TBEFN\cite{TBEFN}, and Zero-DCE++\cite{Zero-DCE++} as a comparison to our method, and we also compared the denoising performance of our method to one of the state-of-the-art denoising methods named FFDNet\cite{ffdnet}. Note that among these methods, DSLR, TBEFN, and FFDNet are supervised methods, while Zero-DCE++, an extension to its original version Zero-DCE\cite{Guo_2020_CVPR}, is trained through unsupervised learning. Note that for DSLR, TBEFN, Zero-DCE++, and FFDNet, we used their corresponding pre-trained versions in all our experiments without fine-tuning. For FFDNet, since it is merely a denoising algorithm, we paired it with a standard gamma correction (GC) with $\gamma$ set to $2.2$.

To see whether performing tone-mapping or denoising first would produce better results, we also trained our models in both configurations. For convenience, we use ``DFTL" (denoise first, tone-map last) to represent the model where tone-mapping is performed after denoising, and ``TFDL" (tone-map first, denoise last) to represent the model where tone-mapping is performed before denoising. The results are summarized in Table \ref{tab:results}, Table \ref{tab:results2}, and Table \ref{tab:results3}. From these quantitative results, we can see that our method outperforms the other methods on most of the benchmarking datasets, and is significantly advantageous on the HDR+ testing set. This is likely due to the fact that, among these datasets we tested, images from the HDR+ testing set contain heavy noise, while images from the LOL dataset and SCIE dataset contain only minimal noise.

\begin{table}[ht!]
\begin{center}
\begin{tabular}{|c|c|c|c|c|}
\hline
Methods & SSIM & PSNR & LPIPS & TMQI\\
\hline\hline
DSLR & 0.500 & 17.358 & 0.412 & \textbf{0.761}\\
\hline
TBEFN & 0.496 & 13.871 & 0.379 & 0.749\\
\hline
Zero-DCE++ & 0.472 & 15.761 & 0.396 & 0.746\\
\hline
FFDNet + GC & 0.526 & 16.989 & 0.459 & 0.680\\
\hline
DFTL (Ours) & \underline{0.668} & \underline{19.905} & \underline{0.349} & \underline{0.759}\\
\hline
TFDL (Ours) & \textbf{0.698} & \textbf{20.168} & \textbf{0.293} & \textbf{0.761}\\
\hline
\end{tabular}
\end{center}
\caption{Quantitative comparison on HDR+ testing set. The best results are printed in boldface. The second best results are underlined.}
\label{tab:results}
\end{table}


\begin{table}[ht!]
\begin{center}
\begin{tabular}{|c|c|c|c|c|}
\hline
Methods & SSIM & PSNR & LPIPS & TMQI\\
\hline\hline
DSLR & \textbf{0.674} & \textbf{14.541} & 0.388 & 0.718\\
\hline
TBEFN & - & - & - & -\\
\hline
Zero-DCE++ & - & - & - & -\\
\hline
FFDNet + GC & \underline{0.542} & \underline{12.721} & 0.475 & 0.686\\
\hline
DFTL (Ours) & 0.517 & 11.727 & \underline{0.352} & \underline{0.767}\\
\hline
TFDL (Ours) & 0.502 & 11.491 & \textbf{0.329} & \textbf{0.778}\\
\hline
\end{tabular}
\end{center}
\caption{Quantitative comparison on the LOL dataset. The best results are printed in boldface. The second best results are underlined. Since TBEFN and Zero-DCE++ are trained on this dataset, we remove them from this comparison.}
\label{tab:results2}
\end{table}

\begin{table}[ht!]
\begin{center}
\begin{tabular}{|c|c|c|c|c|}
\hline
Methods & SSIM & PSNR & LPIPS & TMQI\\
\hline\hline
DSLR & \textbf{0.529} & \underline{13.094} & \underline{0.385} & \textbf{0.706}\\
\hline
TBEFN & - & - & - & -\\
\hline
Zero-DCE++ & - & - & - & -\\
\hline
FFDNet + GC & \underline{0.524} & \textbf{13.656} & 0.450 & 0.531\\
\hline
DFTL (Ours) & \textbf{0.529} & 12.347 & \textbf{0.383} & \underline{0.665}\\
\hline
TFDL (Ours) & 0.517 & 12.210 & \underline{0.385} & 0.646\\
\hline
\end{tabular}
\end{center}
\caption{Quantitative comparison on part 1 of the SCIE dataset. The best results are printed in boldface. The second best results are underlined. Since TBEFN and Zero-DCE++ are trained on this dataset, we remove them from this comparison.}
\label{tab:results3}
\end{table}

\begin{table}[ht!]
\begin{center}
\begin{tabular}{|c|c|c|}
\hline
Methods & MACs(G) & Params(M)\\
\hline\hline
DSLR & 5.876 & 14.931\\
\hline
TBEFN & 3.682 & 0.683\\
\hline
Zero-DCE++ & \textcolor{blue}{0.530} & \textcolor{blue}{0.011}\\
\hline
FFDNet + GC & \textcolor{red}{10.721} & 0.855\\
\hline
DFTL (Ours) & 10.582 & \textcolor{red}{39.806}\\
\hline
TFDL (Ours) & 10.582 & \textcolor{red}{39.806}\\
\hline
\end{tabular}
\end{center}
\caption{Comparison on model sizes and computational complexity. Smallest values are labeled in \textcolor{blue}{blue}, and greatest values are labeled in \textcolor{red}{red}.}
\label{tab:complexity}
\end{table}

\begin{figure*}
\captionsetup[subfigure]{labelformat=empty}
  \centering
    \begin{subfigure}[b]{0.13\textwidth}
    \begin{tikzpicture}[zoomboxarray, zoomboxes below]
        \node [image node] { \includegraphics[width=0.95\textwidth]{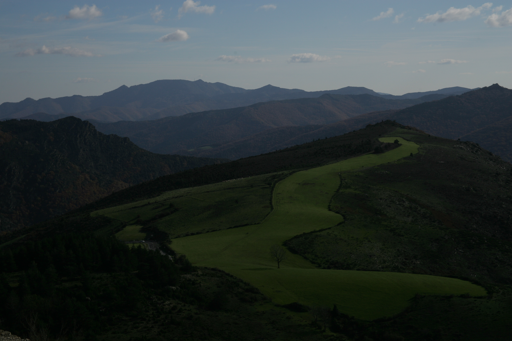} };
        \zoombox[magnification=5, color code=red]{0.175,0.7}
        \zoombox[magnification=5, color code=yellow]{0.61,0.6}
        \zoombox[magnification=5, color code=orange]{0.25,0.29}
        \zoombox[magnification=5,color code=lime]{0.82,0.65}
    \end{tikzpicture}\vspace{-5px}\caption{Input}\vspace{3px} \end{subfigure} \begin{subfigure}[b]{0.13\textwidth}
    \begin{tikzpicture}[zoomboxarray, zoomboxes below]
        \node [image node] { \includegraphics[width=0.95\textwidth]{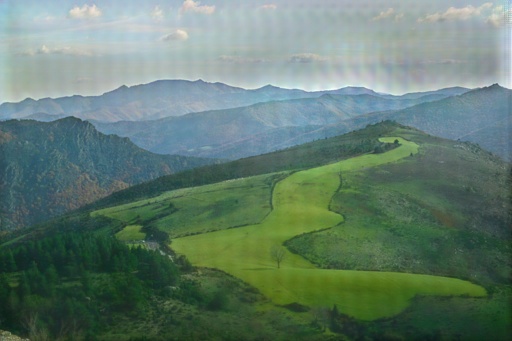} };
        \zoombox[magnification=5, color code=red]{0.175,0.7}
        \zoombox[magnification=5, color code=yellow]{0.61,0.6}
        \zoombox[magnification=5, color code=orange]{0.25,0.29}
        \zoombox[magnification=5,color code=lime]{0.82,0.65}
    \end{tikzpicture}\vspace{-5px}\caption{DSLR}\vspace{3px} \end{subfigure} \begin{subfigure}[b]{0.13\textwidth}
    \begin{tikzpicture}[zoomboxarray, zoomboxes below]
        \node [image node] { \includegraphics[width=0.95\textwidth]{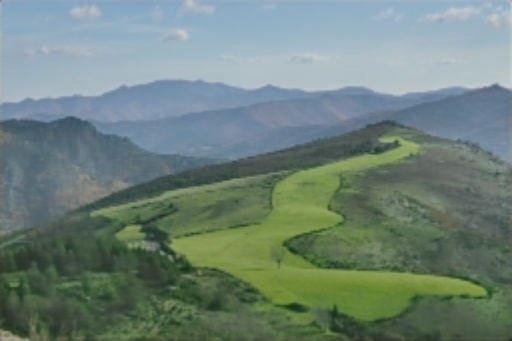} };
        \zoombox[magnification=5, color code=red]{0.175,0.7}
        \zoombox[magnification=5, color code=yellow]{0.61,0.6}
        \zoombox[magnification=5, color code=orange]{0.25,0.29}
        \zoombox[magnification=5,color code=lime]{0.82,0.65}
    \end{tikzpicture}\vspace{-5px}\caption{TBEFN}\vspace{3px} \end{subfigure} \begin{subfigure}[b]{0.13\textwidth}
    \begin{tikzpicture}[zoomboxarray, zoomboxes below]
        \node [image node] { \includegraphics[width=0.95\textwidth]{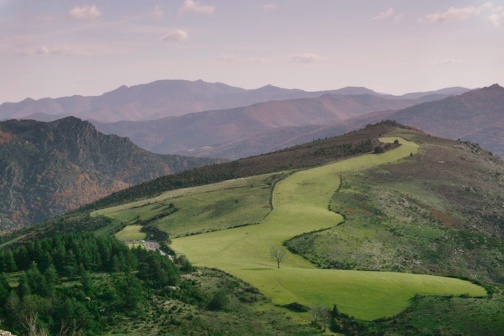} };
        \zoombox[magnification=5, color code=red]{0.175,0.7}
        \zoombox[magnification=5, color code=yellow]{0.61,0.6}
        \zoombox[magnification=5, color code=orange]{0.25,0.29}
        \zoombox[magnification=5,color code=lime]{0.82,0.65}
    \end{tikzpicture}\vspace{-5px}\caption{Zero-DCE++}\vspace{3px} \end{subfigure} \begin{subfigure}[b]{0.13\textwidth}
    \begin{tikzpicture}[zoomboxarray, zoomboxes below]
        \node [image node] { \includegraphics[width=0.95\textwidth]{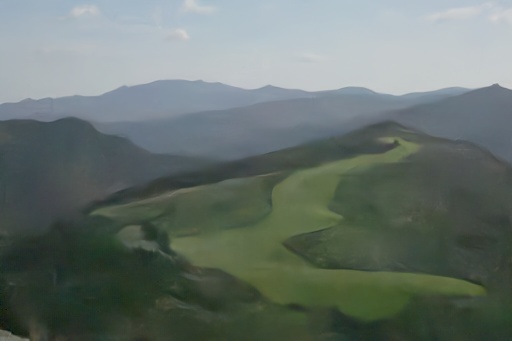} };
        \zoombox[magnification=5, color code=red]{0.175,0.7}
        \zoombox[magnification=5, color code=yellow]{0.61,0.6}
        \zoombox[magnification=5, color code=orange]{0.25,0.29}
        \zoombox[magnification=5,color code=lime]{0.82,0.65}
    \end{tikzpicture}\vspace{-5px}\caption{FFDNet+GC}\vspace{3px} \end{subfigure} \begin{subfigure}[b]{0.13\textwidth}
    \begin{tikzpicture}[zoomboxarray, zoomboxes below]
        \node [image node] { \includegraphics[width=0.95\textwidth]{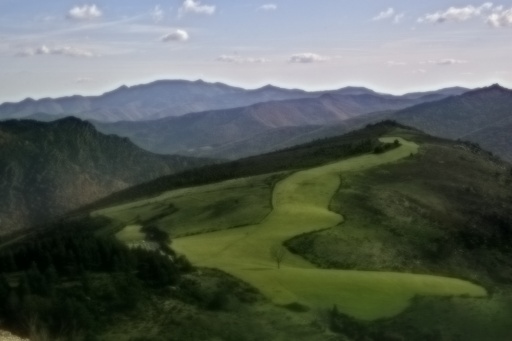} };
        \zoombox[magnification=5, color code=red]{0.175,0.7}
        \zoombox[magnification=5, color code=yellow]{0.61,0.6}
        \zoombox[magnification=5, color code=orange]{0.25,0.29}
        \zoombox[magnification=5,color code=lime]{0.82,0.65}
    \end{tikzpicture}\vspace{-5px}\caption{DFTL (Ours)}\vspace{3px} \end{subfigure} \begin{subfigure}[b]{0.13\textwidth}
    \begin{tikzpicture}[zoomboxarray, zoomboxes below]
        \node [image node] { \includegraphics[width=0.95\textwidth]{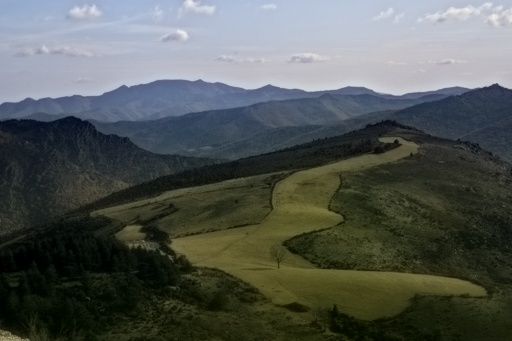} };
        \zoombox[magnification=5, color code=red]{0.175,0.7}
        \zoombox[magnification=5, color code=yellow]{0.61,0.6}
        \zoombox[magnification=5, color code=orange]{0.25,0.29}
        \zoombox[magnification=5,color code=lime]{0.82,0.65}
    \end{tikzpicture}\vspace{-5px}\caption{TFDL (Ours)}\vspace{3px} \end{subfigure} \begin{subfigure}[b]{0.13\textwidth}
    \begin{tikzpicture}[zoomboxarray, zoomboxes below]
        \node [image node] { \includegraphics[width=0.95\textwidth]{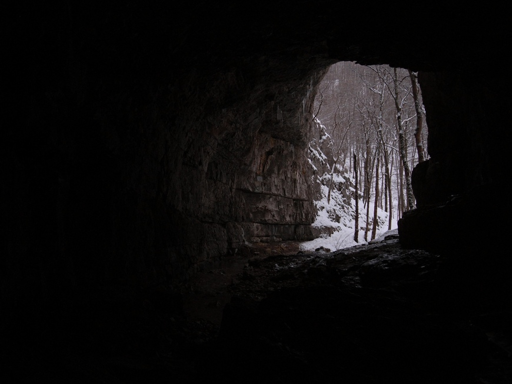} };
        \zoombox[magnification=5, color code=red]{0.5,0.5}
        \zoombox[magnification=5, color code=yellow]{0.51,0.68}
        \zoombox[magnification=5, color code=orange]{0.175,0.51}
        \zoombox[magnification=5,color code=lime]{0.82,0.71}
    \end{tikzpicture}\vspace{-5px}\caption{Input}\vspace{3px} \end{subfigure} \begin{subfigure}[b]{0.13\textwidth}
    \begin{tikzpicture}[zoomboxarray, zoomboxes below]
        \node [image node] { \includegraphics[width=0.95\textwidth]{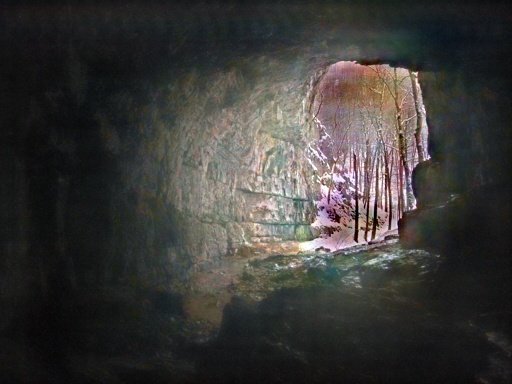} };
        \zoombox[magnification=5, color code=red]{0.5,0.5}
        \zoombox[magnification=5, color code=yellow]{0.51,0.68}
        \zoombox[magnification=5, color code=orange]{0.175,0.51}
        \zoombox[magnification=5,color code=lime]{0.82,0.71}
    \end{tikzpicture}\vspace{-5px}\caption{DSLR}\vspace{3px} \end{subfigure} \begin{subfigure}[b]{0.13\textwidth}
    \begin{tikzpicture}[zoomboxarray, zoomboxes below]
        \node [image node] { \includegraphics[width=0.95\textwidth]{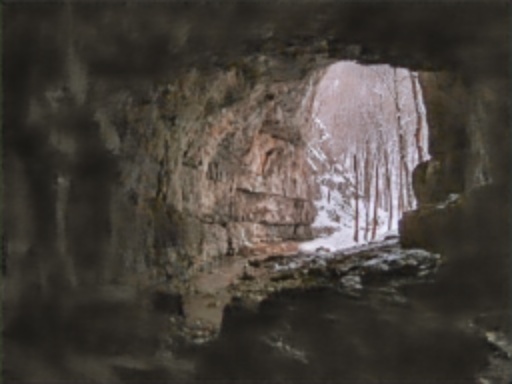} };
        \zoombox[magnification=5, color code=red]{0.5,0.5}
        \zoombox[magnification=5, color code=yellow]{0.51,0.68}
        \zoombox[magnification=5, color code=orange]{0.175,0.51}
        \zoombox[magnification=5,color code=lime]{0.82,0.71}
    \end{tikzpicture}\vspace{-5px}\caption{TBEFN}\vspace{3px} \end{subfigure} \begin{subfigure}[b]{0.13\textwidth}
    \begin{tikzpicture}[zoomboxarray, zoomboxes below]
        \node [image node] { \includegraphics[width=0.95\textwidth]{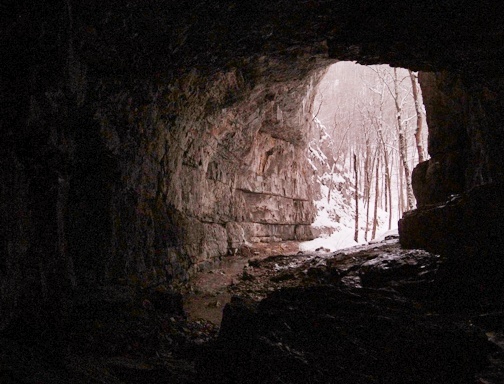} };
        \zoombox[magnification=5, color code=red]{0.5,0.5}
        \zoombox[magnification=5, color code=yellow]{0.51,0.68}
        \zoombox[magnification=5, color code=orange]{0.175,0.51}
        \zoombox[magnification=5,color code=lime]{0.82,0.71}
    \end{tikzpicture}\vspace{-5px}\caption{Zero-DCE++}\vspace{3px} \end{subfigure} \begin{subfigure}[b]{0.13\textwidth}
    \begin{tikzpicture}[zoomboxarray, zoomboxes below]
        \node [image node] { \includegraphics[width=0.95\textwidth]{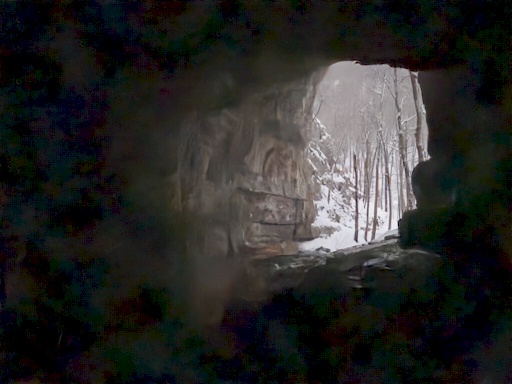} };
        \zoombox[magnification=5, color code=red]{0.5,0.5}
        \zoombox[magnification=5, color code=yellow]{0.51,0.68}
        \zoombox[magnification=5, color code=orange]{0.175,0.51}
        \zoombox[magnification=5,color code=lime]{0.82,0.71}
    \end{tikzpicture}\vspace{-5px}\caption{FFDNet+GC}\vspace{3px} \end{subfigure} \begin{subfigure}[b]{0.13\textwidth}
    \begin{tikzpicture}[zoomboxarray, zoomboxes below]
        \node [image node] { \includegraphics[width=0.95\textwidth]{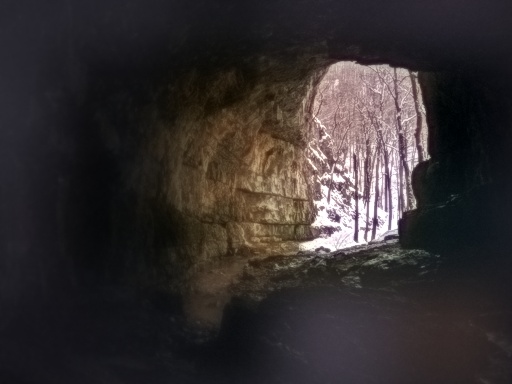} };
        \zoombox[magnification=5, color code=red]{0.5,0.5}
        \zoombox[magnification=5, color code=yellow]{0.51,0.68}
        \zoombox[magnification=5, color code=orange]{0.175,0.51}
        \zoombox[magnification=5,color code=lime]{0.82,0.71}
    \end{tikzpicture}\vspace{-5px}\caption{DFTL (Ours)}\vspace{3px} \end{subfigure} \begin{subfigure}[b]{0.13\textwidth}
    \begin{tikzpicture}[zoomboxarray, zoomboxes below]
        \node [image node] { \includegraphics[width=0.95\textwidth]{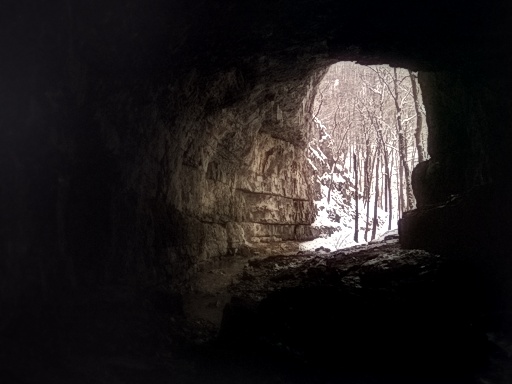} };
        \zoombox[magnification=5, color code=red]{0.5,0.5}
        \zoombox[magnification=5, color code=yellow]{0.51,0.68}
        \zoombox[magnification=5, color code=orange]{0.175,0.51}
        \zoombox[magnification=5,color code=lime]{0.82,0.71}
    \end{tikzpicture}\vspace{-5px}\caption{TFDL (Ours)}\vspace{3px} \end{subfigure}
    \begin{subfigure}[b]{0.13\textwidth}
    \begin{tikzpicture}[zoomboxarray, zoomboxes below]
        \node [image node] { \includegraphics[width=0.95\textwidth]{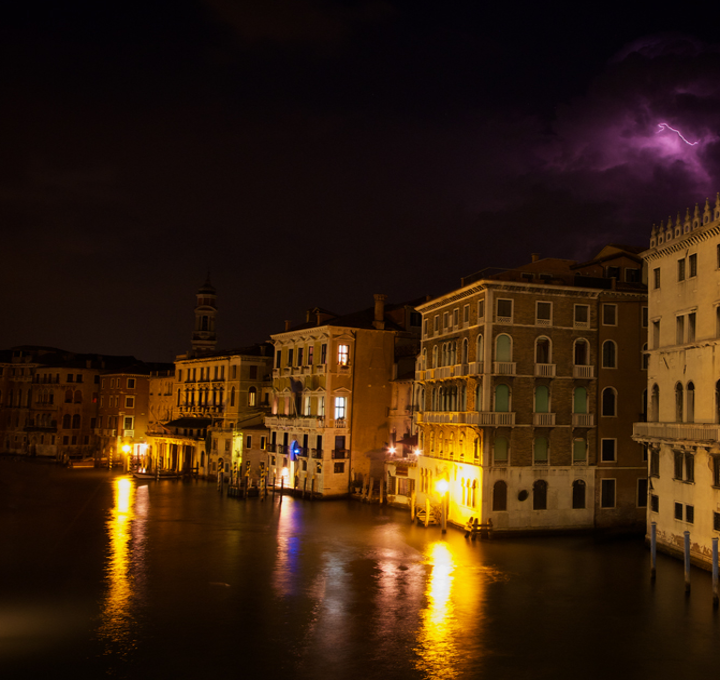} };
        \zoombox[magnification=5, color code=red]{0.2,0.4}
        \zoombox[magnification=5, color code=yellow]{0.7,0.3}
        \zoombox[magnification=5, color code=orange]{0.23,0.17}
        \zoombox[magnification=5,color code=lime]{0.93,0.79}
    \end{tikzpicture}\vspace{-5px}\caption{Input}\vspace{3px} \end{subfigure} \begin{subfigure}[b]{0.13\textwidth}
    \begin{tikzpicture}[zoomboxarray, zoomboxes below]
        \node [image node] { \includegraphics[width=0.95\textwidth]{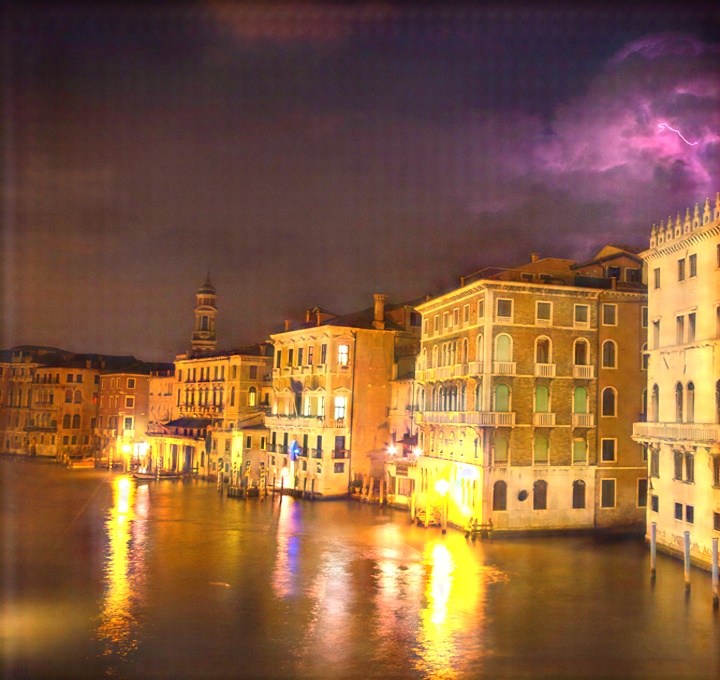} };
        \zoombox[magnification=5, color code=red]{0.2,0.4}
        \zoombox[magnification=5, color code=yellow]{0.7,0.3}
        \zoombox[magnification=5, color code=orange]{0.23,0.17}
        \zoombox[magnification=5,color code=lime]{0.93,0.79}
    \end{tikzpicture}\vspace{-5px}\caption{DSLR}\vspace{3px} \end{subfigure} \begin{subfigure}[b]{0.13\textwidth}
    \begin{tikzpicture}[zoomboxarray, zoomboxes below]
        \node [image node] { \includegraphics[width=0.95\textwidth]{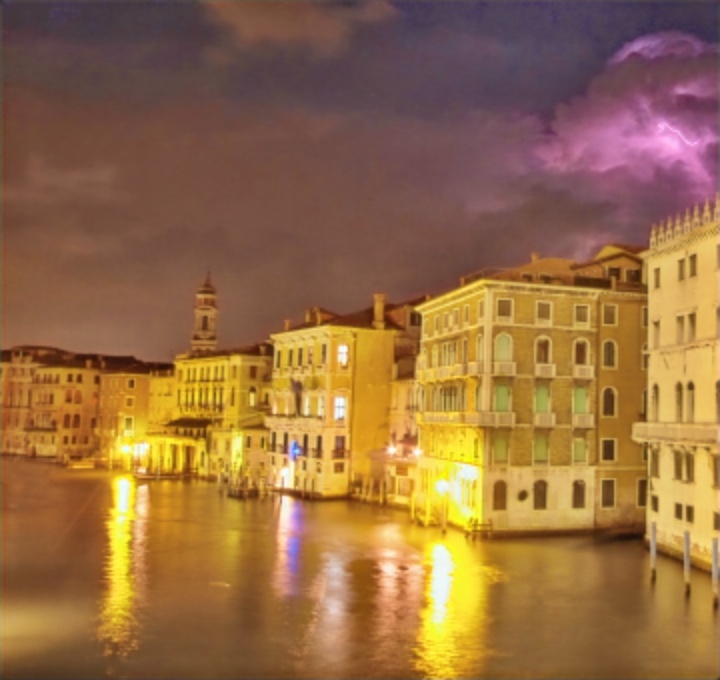} };
        \zoombox[magnification=5, color code=red]{0.2,0.4}
        \zoombox[magnification=5, color code=yellow]{0.7,0.3}
        \zoombox[magnification=5, color code=orange]{0.23,0.17}
        \zoombox[magnification=5,color code=lime]{0.93,0.79}
    \end{tikzpicture}\vspace{-5px}\caption{TBEFN}\vspace{3px} \end{subfigure} \begin{subfigure}[b]{0.13\textwidth}
    \begin{tikzpicture}[zoomboxarray, zoomboxes below]
        \node [image node] { \includegraphics[width=0.95\textwidth]{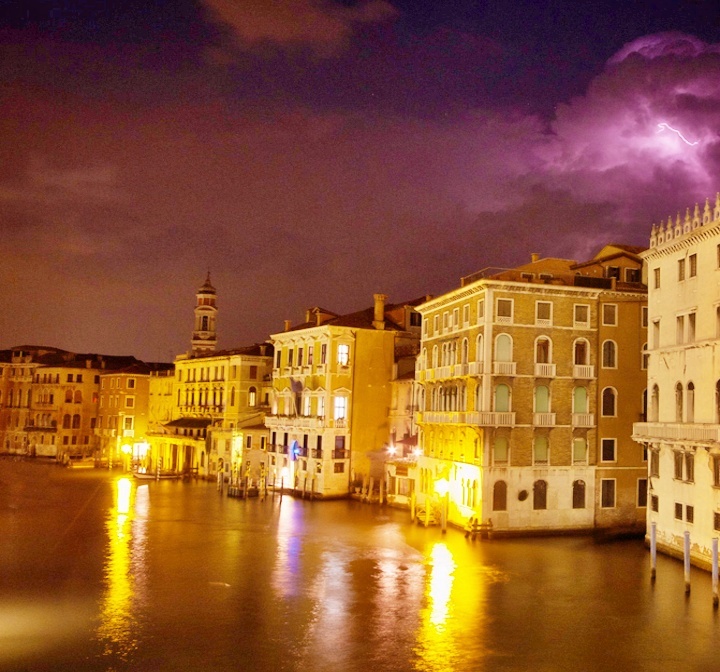} };
        \zoombox[magnification=5, color code=red]{0.2,0.4}
        \zoombox[magnification=5, color code=yellow]{0.7,0.3}
        \zoombox[magnification=5, color code=orange]{0.23,0.17}
        \zoombox[magnification=5,color code=lime]{0.93,0.79}
    \end{tikzpicture}\vspace{-5px}\caption{Zero-DCE++}\vspace{3px} \end{subfigure} \begin{subfigure}[b]{0.13\textwidth}
    \begin{tikzpicture}[zoomboxarray, zoomboxes below]
        \node [image node] { \includegraphics[width=0.95\textwidth]{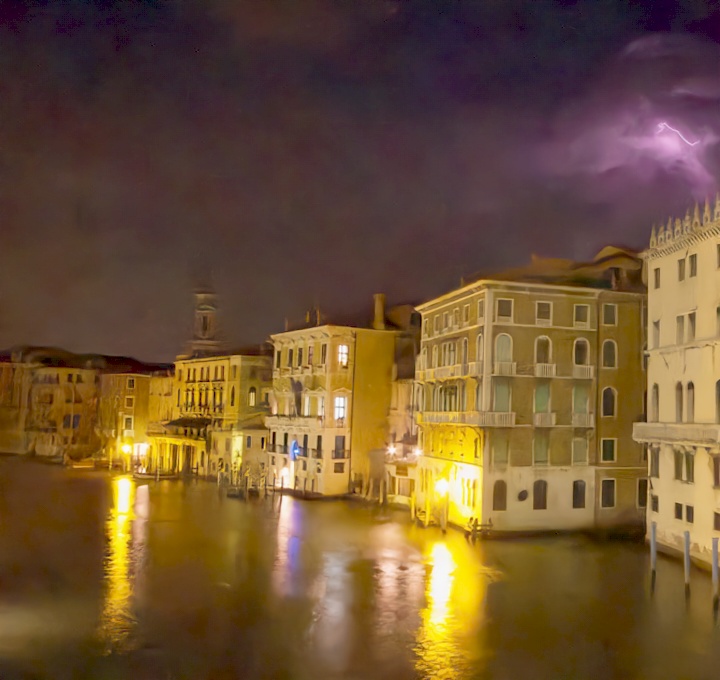} };
        \zoombox[magnification=5, color code=red]{0.2,0.4}
        \zoombox[magnification=5, color code=yellow]{0.7,0.3}
        \zoombox[magnification=5, color code=orange]{0.23,0.17}
        \zoombox[magnification=5,color code=lime]{0.93,0.79}
    \end{tikzpicture}\vspace{-5px}\caption{FFDNet+GC}\vspace{3px} \end{subfigure} \begin{subfigure}[b]{0.13\textwidth}
    \begin{tikzpicture}[zoomboxarray, zoomboxes below]
        \node [image node] { \includegraphics[width=0.95\textwidth]{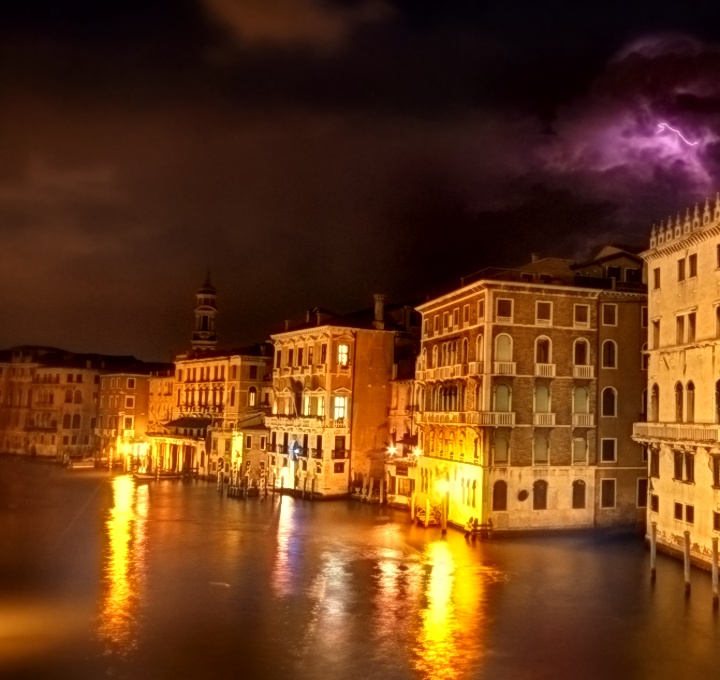} };
        \zoombox[magnification=5, color code=red]{0.2,0.4}
        \zoombox[magnification=5, color code=yellow]{0.7,0.3}
        \zoombox[magnification=5, color code=orange]{0.23,0.17}
        \zoombox[magnification=5,color code=lime]{0.93,0.79}
    \end{tikzpicture}\vspace{-5px}\caption{DFTL (Ours)}\vspace{3px} \end{subfigure} \begin{subfigure}[b]{0.13\textwidth}
    \begin{tikzpicture}[zoomboxarray, zoomboxes below]
        \node [image node] { \includegraphics[width=0.95\textwidth]{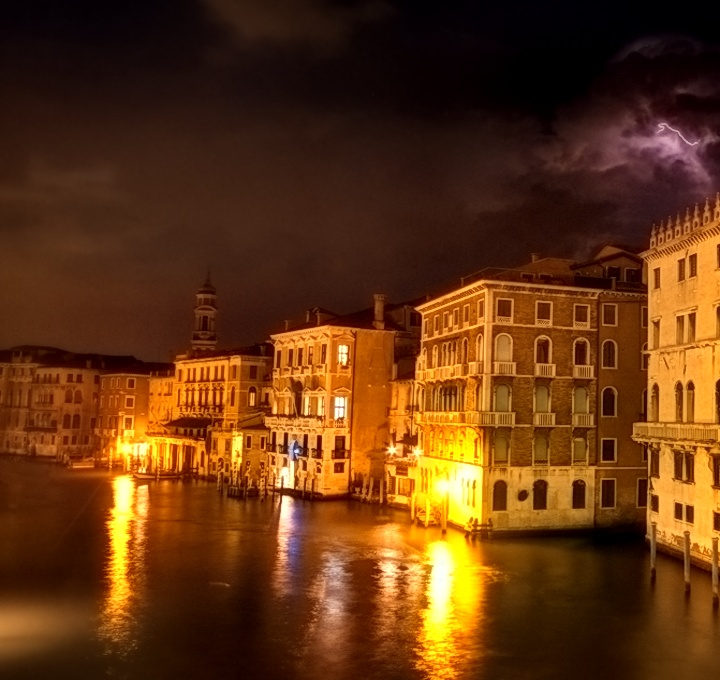} };
        \zoombox[magnification=5, color code=red]{0.2,0.4}
        \zoombox[magnification=5, color code=yellow]{0.7,0.3}
        \zoombox[magnification=5, color code=orange]{0.23,0.17}
        \zoombox[magnification=5,color code=lime]{0.93,0.79}
    \end{tikzpicture}\vspace{-5px}\caption{TFDL (Ours)}\vspace{3px} \end{subfigure} \begin{subfigure}[b]{0.13\textwidth}
    \begin{tikzpicture}[zoomboxarray, zoomboxes below]
        \node [image node] { \includegraphics[width=0.95\textwidth]{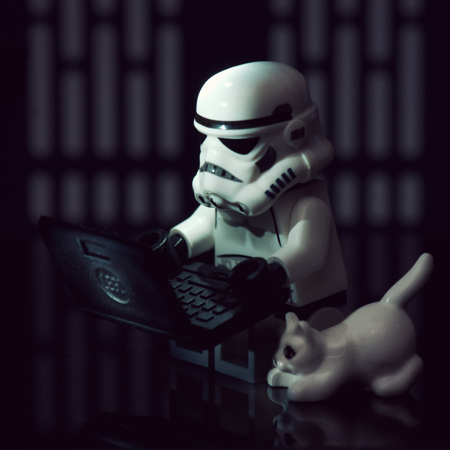} };
        \zoombox[magnification=5, color code=red]{0.5,0.3}
        \zoombox[magnification=5, color code=yellow]{0.51,0.62}
        \zoombox[magnification=5, color code=orange]{0.2,0.63}
        \zoombox[magnification=5,color code=lime]{0.82,0.52}
    \end{tikzpicture}\vspace{-5px}\caption{Input}\vspace{3px} \end{subfigure} \begin{subfigure}[b]{0.13\textwidth}
    \begin{tikzpicture}[zoomboxarray, zoomboxes below]
        \node [image node] { \includegraphics[width=0.95\textwidth]{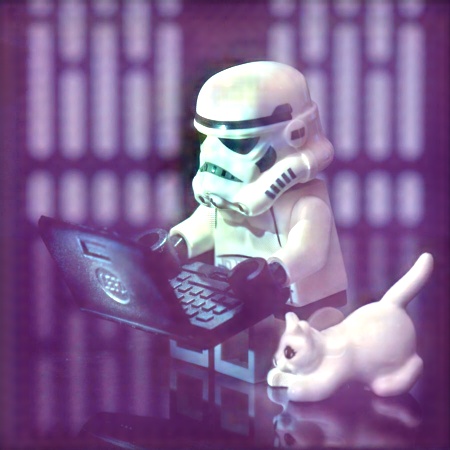} };
        \zoombox[magnification=5, color code=red]{0.5,0.3}
        \zoombox[magnification=5, color code=yellow]{0.51,0.62}
        \zoombox[magnification=5, color code=orange]{0.2,0.63}
        \zoombox[magnification=5,color code=lime]{0.82,0.52}
    \end{tikzpicture}\vspace{-5px}\caption{DSLR}\vspace{3px} \end{subfigure} \begin{subfigure}[b]{0.13\textwidth}
    \begin{tikzpicture}[zoomboxarray, zoomboxes below]
        \node [image node] { \includegraphics[width=0.95\textwidth]{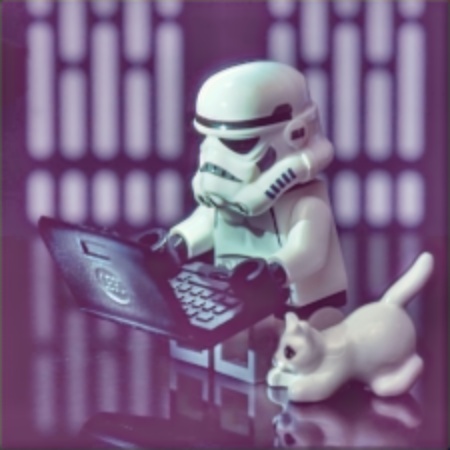} };
        \zoombox[magnification=5, color code=red]{0.5,0.3}
        \zoombox[magnification=5, color code=yellow]{0.51,0.62}
        \zoombox[magnification=5, color code=orange]{0.2,0.63}
        \zoombox[magnification=5,color code=lime]{0.82,0.52}
    \end{tikzpicture}\vspace{-5px}\caption{TBEFN}\vspace{3px} \end{subfigure} \begin{subfigure}[b]{0.13\textwidth}
    \begin{tikzpicture}[zoomboxarray, zoomboxes below]
        \node [image node] { \includegraphics[width=0.95\textwidth]{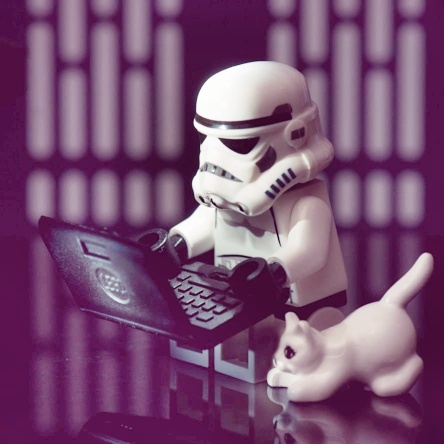} };
        \zoombox[magnification=5, color code=red]{0.5,0.3}
        \zoombox[magnification=5, color code=yellow]{0.51,0.62}
        \zoombox[magnification=5, color code=orange]{0.2,0.63}
        \zoombox[magnification=5,color code=lime]{0.82,0.52}
    \end{tikzpicture}\vspace{-5px}\caption{Zero-DCE++}\vspace{3px} \end{subfigure} \begin{subfigure}[b]{0.13\textwidth}
    \begin{tikzpicture}[zoomboxarray, zoomboxes below]
        \node [image node] { \includegraphics[width=0.95\textwidth]{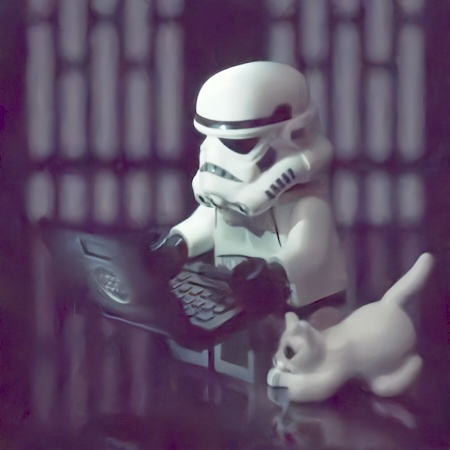} };
        \zoombox[magnification=5, color code=red]{0.5,0.3}
        \zoombox[magnification=5, color code=yellow]{0.51,0.62}
        \zoombox[magnification=5, color code=orange]{0.2,0.63}
        \zoombox[magnification=5,color code=lime]{0.82,0.52}
    \end{tikzpicture}\vspace{-5px}\caption{FFDNet+GC}\vspace{3px} \end{subfigure} \begin{subfigure}[b]{0.13\textwidth}
    \begin{tikzpicture}[zoomboxarray, zoomboxes below]
        \node [image node] { \includegraphics[width=0.95\textwidth]{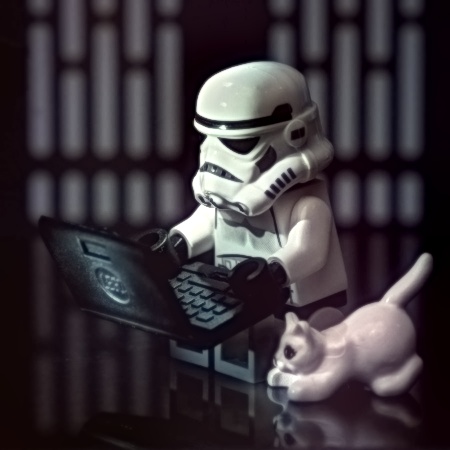} };
        \zoombox[magnification=5, color code=red]{0.5,0.3}
        \zoombox[magnification=5, color code=yellow]{0.51,0.62}
        \zoombox[magnification=5, color code=orange]{0.2,0.63}
        \zoombox[magnification=5,color code=lime]{0.82,0.52}
    \end{tikzpicture}\vspace{-5px}\caption{DFTL (Ours)}\vspace{3px} \end{subfigure} \begin{subfigure}[b]{0.13\textwidth}
    \begin{tikzpicture}[zoomboxarray, zoomboxes below]
        \node [image node] { \includegraphics[width=0.95\textwidth]{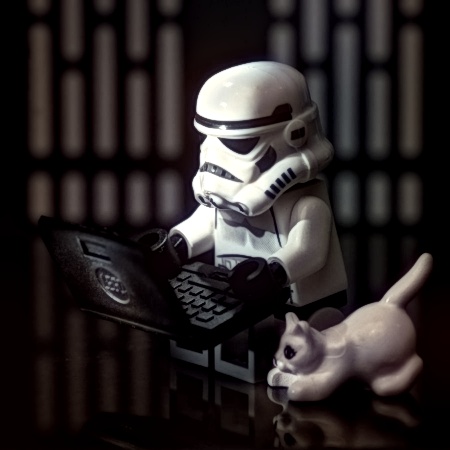} };
        \zoombox[magnification=5, color code=red]{0.5,0.3}
        \zoombox[magnification=5, color code=yellow]{0.51,0.62}
        \zoombox[magnification=5, color code=orange]{0.2,0.63}
        \zoombox[magnification=5,color code=lime]{0.82,0.52}
    \end{tikzpicture}\vspace{-5px}\caption{TFDL (Ours)}\vspace{3px} \end{subfigure}
  \caption{Qualitative comparisons between methods on examples from the MEF dataset (top two rows) and LIME dataset (bottom two rows). Four sub-regions selected from each example are zoomed in 5 times to show more details.}
\label{fig:example2}
\end{figure*}


For qualitative comparison, examples from the HDR+ testing set with zoomed-in views are shown in Figure \ref{fig:example}. We also show examples from the LOL dataset and part 1 of the SCIE dataset in Figure \ref{fig:example1} to demonstrate the effectiveness of our methods on an unseen dataset. From these representative examples, we can see that both of our methods outperform the other methods in the comparison in terms of both tone-mapping and denoising. Compared to TBEFN and Zero-DCE++, DSLR does a better job of tone-mapping the image to a natural brightness level, especially in the extreme low-light scenes. However, all these three methods suffer from the amount of noise in the input. In terms of denoising, FFDNet does a pretty good job at removing heavy noise in the input, but it may overly blur the image and smooth out important details or high-frequency information. Our methods significantly outperform other methods on both the HDR+ dataset and the LOL dataset, in terms of both the tone-mapping quality and denoising quality. To be specific, our methods can tone-map the low-light inputs to appropriate brightness levels while maintaining natural contrasts. Moreover, our methods effectively remove the noise while keeping most of the important details intact. 

\subsubsection{No-Reference Comparison}

To further showcase the effectiveness of our methods on unseen datasets, we also tested our models on two more popular datasets, namely the MEF dataset \cite{mef} and the LIME dataset \cite{lime}. Since these datasets do not contain paired ground-truth images, we only show outputs from our models as well as the other methods in Figure \ref{fig:example2} for a qualitative comparison. From these no-reference results, we can see that our results have better contrast, look more natural, and are aesthetically more pleasing, compared to the results from other methods. 

\subsubsection{Computational Complexity Comparison}

In order to compare the efficiency of these methods, we also measured the model size and computational complexity of each model and summarized the results in Table \ref{tab:complexity}. To be specific, we measured the number of MAC (multiply-accumulate operations) as well as learnable parameters in these models with a 3-channel input image in size $224\times 224$. We can see from the table that, although our joint framework performs well in both the quantitative and qualitative comparisons, it does have a trade-off in its model size. However, considering that our method performs both tone-mapping and denoising at the same time and that denoising algorithms are often computationally intensive, this trade-off should be worthy. 



\section{Conclusions}

In this paper, we have proposed a composite HDR image enhancement model that jointly performs HDR image tone-mapping and image denoising in a multi-scale framework. Based on the testing results on the HDR+ dataset shown in the previous section, our joint multi-scale tone mapping and denoising model for HDR image enhancement outperforms other state-of-the-art image enhancement methods both quantitatively and qualitatively. Our experiments on the ordering of the tone-mapping operator (TMO) and the denoising operator demonstrate that applying tone mapping before denoising yields slightly better results in our joint image enhancement framework. In the future, we plan to exploit the combination of multi-image tone-mapping and multi-image denoising for HDR images.

{\small
\bibliographystyle{ieee_fullname}
\bibliography{egbib}

\begin{thebibliography}{10}\itemsep=-1pt

\bibitem{afifi2021learning}
Mahmoud Afifi, Konstantinos~G Derpanis, Bj{\"o}rn Ommer, and Michael~S Brown.
\newblock Learning multi-scale photo exposure correction.
\newblock In {\em Proceedings of the IEEE Conference on Computer Vision and
  Pattern Recognition}, 2021.

\bibitem{Anwar_2020_CVPR_Workshops}
Saeed Anwar, Cong~Phuoc Huynh, and Fatih Porikli.
\newblock Identity enhanced residual image denoising.
\newblock In {\em Proceedings of the IEEE/CVF Conference on Computer Vision and
  Pattern Recognition (CVPR) Workshops}, June 2020.

\bibitem{Bao_2020_CVPR_Workshops}
Long Bao, Zengli Yang, Shuangquan Wang, Dongwoon Bai, and Jungwon Lee.
\newblock Real image denoising based on multi-scale residual dense block and
  cascaded u-net with block-connection.
\newblock In {\em Proceedings of the IEEE/CVF Conference on Computer Vision and
  Pattern Recognition (CVPR) Workshops}, June 2020.

\bibitem{Burger2016}
Wilhelm Burger and Mark~J. Burge.
\newblock {\em Scale-Invariant Feature Transform (SIFT)}, pages 609--664.
\newblock Springer London, London, 2016.

\bibitem{Cai2018deep}
Jianrui Cai, Shuhang Gu, and Lei Zhang.
\newblock Learning a deep single image contrast enhancer from multi-exposure
  images.
\newblock {\em IEEE Transactions on Image Processing}, 27(4):2049--2062, 2018.

\bibitem{Chen_2019_ICCV}
Chen Chen, Qifeng Chen, Minh~N. Do, and Vladlen Koltun.
\newblock Seeing motion in the dark.
\newblock In {\em Proceedings of the IEEE/CVF International Conference on
  Computer Vision (ICCV)}, October 2019.

\bibitem{Chen_2018_CVPR}
Chen Chen, Qifeng Chen, Jia Xu, and Vladlen Koltun.
\newblock Learning to see in the dark.
\newblock In {\em Proceedings of the IEEE Conference on Computer Vision and
  Pattern Recognition (CVPR)}, June 2018.

\bibitem{Chen_gan}
Jingwen Chen, Jiawei Chen, Hongyang Chao, and Ming Yang.
\newblock Image blind denoising with generative adversarial network based noise
  modeling.
\newblock In {\em 2018 IEEE/CVF Conference on Computer Vision and Pattern
  Recognition}, pages 3155--3164, 2018.

\bibitem{10.5555/2969239.2969405}
Emily Denton, Soumith Chintala, Arthur Szlam, and Rob Fergus.
\newblock Deep generative image models using a laplacian pyramid of adversarial
  networks.
\newblock In {\em Proceedings of the 28th International Conference on Neural
  Information Processing Systems - Volume 1}, NIPS'15, page 1486–1494,
  Cambridge, MA, USA, 2015. MIT Press.

\bibitem{durand2002fast}
Fr{\'e}do Durand and Julie Dorsey.
\newblock Fast bilateral filtering for the display of high-dynamic-range
  images.
\newblock In {\em Proceedings of the 29th annual conference on Computer
  graphics and interactive techniques}, pages 257--266, 2002.

\bibitem{thibaud}
Thibaud Ehret, Axel Davy, Pablo Arias, and Gabriele Facciolo.
\newblock Joint demosaicing and denoising by overfitting of bursts of raw
  images.
\newblock {\em CoRR}, abs/1905.05092, 2019.

\bibitem{Fan}
Minhao Fan, Wenjing Wang, Wenhan Yang, and Jiaying Liu.
\newblock Integrating semantic segmentation and retinex model for low-light
  image enhancement.
\newblock In {\em Proceedings of the 28th ACM International Conference on
  Multimedia}, MM '20, page 2317–2325, New York, NY, USA, 2020. Association
  for Computing Machinery.

\bibitem{joint_dm}
Micha\"{e}l Gharbi, Gaurav Chaurasia, Sylvain Paris, and Fr\'{e}do Durand.
\newblock Deep joint demosaicking and denoising.
\newblock {\em ACM Trans. Graph.}, 35(6), Nov. 2016.

\bibitem{burst_denoise}
Cl{\'{e}}ment Godard, Kevin Matzen, and Matt Uyttendaele.
\newblock Deep burst denoising.
\newblock {\em CoRR}, abs/1712.05790, 2017.

\bibitem{noise_estimation}
Miguel Granados, Tun\c{c}~Ozan Ayd\i{}n, J.~Rafael Tena, Jean-Fran\c{c}ois
  Lalonde, and Christian Theobalt.
\newblock {HDR} image noise estimation for denoising tone mapped images.
\newblock In {\em Proceedings of the 12th European Conference on Visual Media
  Production}, CVMP '15, New York, NY, USA, 2015. Association for Computing
  Machinery.

\bibitem{Guo_2020_CVPR}
Chunle Guo, Chongyi Li, Jichang Guo, Chen~Change Loy, Junhui Hou, Sam Kwong,
  and Runmin Cong.
\newblock Zero-reference deep curve estimation for low-light image enhancement.
\newblock In {\em Proceedings of the IEEE/CVF Conference on Computer Vision and
  Pattern Recognition (CVPR)}, June 2020.

\bibitem{Guo2019Cbdnet}
Shi Guo, Zifei Yan, Kai Zhang, Wangmeng Zuo, and Lei Zhang.
\newblock Toward convolutional blind denoising of real photographs.
\newblock {\em 2019 IEEE Conference on Computer Vision and Pattern Recognition
  (CVPR)}, 2019.

\bibitem{lime}
Xiaojie Guo, Yu Li, and Haibin Ling.
\newblock {LIME}: Low-light image enhancement via illumination map estimation.
\newblock {\em IEEE Transactions on Image Processing}, 26(2):982--993, 2017.

\bibitem{hasinoff2016burst}
Samuel~W. Hasinoff, Dillon Sharlet, Ryan Geiss, Andrew Adams, Jonathan~T.
  Barron, Florian Kainz, Jiawen Chen, and Marc Levoy.
\newblock Burst photography for high dynamic range and low-light imaging on
  mobile cameras.
\newblock {\em ACM Transactions on Graphics (Proc. SIGGRAPH Asia)}, 35(6),
  2016.

\bibitem{he2020conditional}
Jingwen He, Yihao Liu, Yu Qiao, and Chao Dong.
\newblock Conditional sequential modulation for efficient global image
  retouching.
\newblock {\em arXiv preprint arXiv:2009.10390}, 2020.

\bibitem{hu2019frame}
Litao Hu, Jan Allebach, Gautam Glowala, Sathya Sundaram, and Perry Lee.
\newblock Frame detection for photos of online fashion items.
\newblock In {\em Proc. IS\&T Int'l. Symp. on Electronic Imaging: Imaging and
  Multimedia Analytics in a Web and Mobile World}, pages 412--1--412--5, 2019.

\bibitem{Hu2020DocumentIQ}
Litao Hu, Zhenhua Hu, P{\'e}ter Bauer, Todd Harris, and Jan~P. Allebach.
\newblock Document image quality assessment with relaying reference to
  determine minimum readable resolution for compression.
\newblock {\em Electronic Imaging}, 2020.

\bibitem{hu2021deep}
Litao Hu, Zhenhua Hu, Peter Bauer, Todd~J. Harris, and Jan~P. Allebach.
\newblock Deep learning approaches to determining optimal resolution for
  scanned text documents.
\newblock In {\em Proc. IS\&T Int'l. Symp. on Electronic Imaging: Color Imaging
  XXVI: Displaying, Processing, Hardcopy, and Applications}, pages
  243--1--243--8, 2021.

\bibitem{Hu2019NonnativeCD}
Litao Hu, Karthick Shankar, Zhi Li, Zhen Yuan, Jan~P. Allebach, Gautam Glowala,
  Sathya Sundaram, and Perry Lee.
\newblock Non-native contents detection and localization for online fashion
  images.
\newblock {\em Electronic Imaging}, 2019.

\bibitem{hu2020relation}
Zhenhua Hu, Litao Hu, Peter Bauer, Todd Harris, and Jan Allebach.
\newblock Relation between image quality and scan resolution: Part i.
\newblock In {\em Proc. IS\&T Int'l. Symp. on Electronic Imaging: Image Quality
  and System Performance XVII}, pages 322--1--322--7, 2020.

\bibitem{Jiang_2019_ICCV}
Haiyang Jiang and Yinqiang Zheng.
\newblock Learning to see moving objects in the dark.
\newblock In {\em Proceedings of the IEEE/CVF International Conference on
  Computer Vision (ICCV)}, October 2019.

\bibitem{EnlightenGAN}
Yifan Jiang, Xinyu Gong, Ding Liu, Yu Cheng, Chen Fang, Xiaohui Shen, Jianchao
  Yang, Pan Zhou, and Zhangyang Wang.
\newblock Enlightengan: Deep light enhancement without paired supervision.
\newblock {\em IEEE Transactions on Image Processing}, 30:2340--2349, 2021.

\bibitem{filippos}
Filippos Kokkinos and Stamatios Lefkimmiatis.
\newblock Deep image demosaicking using a cascade of convolutional residual
  denoising networks.
\newblock {\em CoRR}, abs/1803.05215, 2018.

\bibitem{8100101}
Wei-Sheng Lai, Jia-Bin Huang, Narendra Ahuja, and Ming-Hsuan Yang.
\newblock Deep laplacian pyramid networks for fast and accurate
  super-resolution.
\newblock In {\em 2017 IEEE Conference on Computer Vision and Pattern
  Recognition (CVPR)}, pages 5835--5843, 2017.

\bibitem{pmlr-v80-lehtinen18a}
Jaakko Lehtinen, Jacob Munkberg, Jon Hasselgren, Samuli Laine, Tero Karras,
  Miika Aittala, and Timo Aila.
\newblock {N}oise2{N}oise: Learning image restoration without clean data.
\newblock In Jennifer Dy and Andreas Krause, editors, {\em Proceedings of the
  35th International Conference on Machine Learning}, volume~80 of {\em
  Proceedings of Machine Learning Research}, pages 2965--2974. PMLR, 10--15 Jul
  2018.

\bibitem{Zero-DCE++}
Chongyi Li, Chunle Guo, and Chen~Change Loy.
\newblock Learning to enhance low-light image via zero-reference deep curve
  estimation.
\newblock In {\em IEEE Transactions on Pattern Analysis and Machine
  Intelligence}, 2021.

\bibitem{LI201815}
Chongyi Li, Jichang Guo, Fatih Porikli, and Yanwei Pang.
\newblock Lightennet: A convolutional neural network for weakly illuminated
  image enhancement.
\newblock {\em Pattern Recognition Letters}, 104:15--22, 2018.

\bibitem{DSLR}
Seokjae Lim and Wonjun Kim.
\newblock {DSLR}: Deep stacked {Laplacian} restorer for low-light image
  enhancement.
\newblock {\em IEEE Transactions on Multimedia}, pages 1--1, 2020.

\bibitem{Liu_2020_CVPR}
Lin Liu, Xu Jia, Jianzhuang Liu, and Qi Tian.
\newblock Joint demosaicing and denoising with self guidance.
\newblock In {\em Proceedings of the IEEE/CVF Conference on Computer Vision and
  Pattern Recognition (CVPR)}, June 2020.

\bibitem{Liu_2020_CVPR_Workshops}
Wei Liu, Qiong Yan, and Yuzhi Zhao.
\newblock Densely self-guided wavelet network for image denoising.
\newblock In {\em Proceedings of the IEEE/CVF Conference on Computer Vision and
  Pattern Recognition (CVPR) Workshops}, June 2020.

\bibitem{GradNet}
Yang Liu, Saeed Anwar, Liang Zheng, and Qi Tian.
\newblock {GradNet} image denoising.
\newblock In {\em Proceedings of the IEEE/CVF Conference on Computer Vision and
  Pattern Recognition (CVPR) Workshops}, June 2020.

\bibitem{TBEFN}
Kun Lu and Lihong Zhang.
\newblock {TBEFN}: A two-branch exposure-fusion network for low-light image
  enhancement.
\newblock {\em IEEE Transactions on Multimedia}, pages 1--1, 2020.

\bibitem{Lv}
Feifan Lv, Bo Liu, and Feng Lu.
\newblock Fast enhancement for non-uniform illumination images using
  light-weight {CNNs}.
\newblock In {\em Proceedings of the 28th ACM International Conference on
  Multimedia}, MM '20, page 1450–1458, New York, NY, USA, 2020. Association
  for Computing Machinery.

\bibitem{mef}
Kede Ma, Kai Zeng, and Zhou Wang.
\newblock Perceptual quality assessment for multi-exposure image fusion.
\newblock {\em IEEE Transactions on Image Processing}, 24(11):3345--3356, 2015.

\bibitem{rcf}
Antoine Monod, Julie Delon, and Thomas Veit.
\newblock {An Analysis and Implementation of the HDR+ Burst Denoising Method}.
\newblock {\em {Image Processing On Line}}, 11:142--169, 2021.
\newblock \url{https://doi.org/10.5201/ipol.2021.336}.

\bibitem{multiscale_DCT}
Nicola Pierazzo, Jean-Michel Morel, and Gabriele Facciolo.
\newblock {Multi-Scale DCT Denoising}.
\newblock {\em {Image Processing On Line}}, 7:288--308, 2017.
\newblock \url{https://doi.org/10.5201/ipol.2017.201}.

\bibitem{Ren}
Wenqi Ren, Sifei Liu, Lin Ma, Qianqian Xu, Xiangyu Xu, Xiaochun Cao, Junping
  Du, and Ming-Hsuan Yang.
\newblock Low-light image enhancement via a deep hybrid network.
\newblock {\em IEEE Transactions on Image Processing}, 28(9):4364--4375, 2019.

\bibitem{denoise_survey}
Chunwei Tian, Yong Xu, Lunke Fei, and Ke Yan.
\newblock Deep learning for image denoising: A survey.
\newblock In Jeng-Shyang Pan, Jerry Chun-Wei Lin, Bixia Sui, and Shih-Pang
  Tseng, editors, {\em Genetic and Evolutionary Computing}, pages 563--572,
  Singapore, 2019. Springer Singapore.

\bibitem{SIDGAN}
Danai Triantafyllidou, Sean Moran, Steven McDonagh, Sarah Parisot, and Gregory
  Slabaugh.
\newblock Low light video enhancement using synthetic data produced with an
  intermediate domain mapping.
\newblock In Andrea Vedaldi, Horst Bischof, Thomas Brox, and Jan-Michael Frahm,
  editors, {\em Computer Vision -- ECCV 2020}, pages 103--119, Cham, 2020.
  Springer International Publishing.

\bibitem{Vaksman_2020_CVPR_Workshops}
Gregory Vaksman, Michael Elad, and Peyman Milanfar.
\newblock Lidia: Lightweight learned image denoising with instance adaptation.
\newblock In {\em Proceedings of the IEEE/CVF Conference on Computer Vision and
  Pattern Recognition (CVPR) Workshops}, June 2020.

\bibitem{DLN}
Li-Wen Wang, Zhi-Song Liu, Wan-Chi Siu, and Daniel P.~K. Lun.
\newblock Lightening network for low-light image enhancement.
\newblock {\em IEEE Transactions on Image Processing}, 29:7984--7996, 2020.

\bibitem{Wang_2019_CVPR}
Ruixing Wang, Qing Zhang, Chi-Wing Fu, Xiaoyong Shen, Wei-Shi Zheng, and Jiaya
  Jia.
\newblock Underexposed photo enhancement using deep illumination estimation.
\newblock In {\em Proceedings of the IEEE/CVF Conference on Computer Vision and
  Pattern Recognition (CVPR)}, June 2019.

\bibitem{Wang}
Yang Wang, Yang Cao, Zheng-Jun Zha, Jing Zhang, Zhiwei Xiong, Wei Zhang, and
  Feng Wu.
\newblock Progressive retinex: Mutually reinforced illumination-noise
  perception network for low-light image enhancement.
\newblock In {\em Proceedings of the 27th ACM International Conference on
  Multimedia}, MM '19, page 2015–2023, New York, NY, USA, 2019. Association
  for Computing Machinery.

\bibitem{Chen2018Retinex}
Chen Wei, Wenjing Wang, Wenhan Yang, and Jiaying Liu.
\newblock Deep retinex decomposition for low-light enhancement.
\newblock In {\em British Machine Vision Conference}, 2018.

\bibitem{Wei2018DeepRD}
Chen Wei, Wenjing Wang, Wenhan Yang, and Jiaying Liu.
\newblock Deep retinex decomposition for low-light enhancement.
\newblock In {\em BMVC}, 2018.

\bibitem{Xu_2020_CVPR}
Ke Xu, Xin Yang, Baocai Yin, and Rynson~W.H. Lau.
\newblock Learning to restore low-light images via
  decomposition-and-enhancement.
\newblock In {\em Proceedings of the IEEE/CVF Conference on Computer Vision and
  Pattern Recognition (CVPR)}, June 2020.

\bibitem{Yang_2020_CVPR}
Wenhan Yang, Shiqi Wang, Yuming Fang, Yue Wang, and Jiaying Liu.
\newblock From fidelity to perceptual quality: A semi-supervised approach for
  low-light image enhancement.
\newblock In {\em Proceedings of the IEEE/CVF Conference on Computer Vision and
  Pattern Recognition (CVPR)}, June 2020.

\bibitem{tmqi}
Hojatollah Yeganeh and Zhou Wang.
\newblock Objective quality assessment of tone-mapped images.
\newblock {\em IEEE Transactions on Image Processing}, 22(2):657--667, 2013.

\bibitem{DCT_denoise}
Guoshen Yu and Guillermo Sapiro.
\newblock {DCT image denoising: a simple and effective image denoising
  algorithm}.
\newblock {\em {Image Processing On Line}}, 1, 2011.

\bibitem{ducnn}
Kai Zhang, Wangmeng Zuo, Yunjin Chen, Deyu Meng, and Lei Zhang.
\newblock Beyond a {Gaussian} denoiser: Residual learning of deep {CNN} for
  image denoising.
\newblock {\em IEEE Transactions on Image Processing}, 26(7):3142--3155, 2017.

\bibitem{ffdnet}
Kai Zhang, Wangmeng Zuo, and Lei Zhang.
\newblock {FFDNet}: Toward a fast and flexible solution for {CNN}-based image
  denoising.
\newblock {\em IEEE Transactions on Image Processing}, 27(9):4608--4622, 2018.

\bibitem{ExCNet}
Lin Zhang, Lijun Zhang, Xiao Liu, Ying Shen, Shaoming Zhang, and Shengjie Zhao.
\newblock Zero-shot restoration of back-lit images using deep internal
  learning.
\newblock In {\em Proceedings of the 27th ACM International Conference on
  Multimedia}, MM '19, page 1623–1631, New York, NY, USA, 2019. Association
  for Computing Machinery.

\bibitem{zhang2018perceptual}
Richard Zhang, Phillip Isola, Alexei~A Efros, Eli Shechtman, and Oliver Wang.
\newblock The unreasonable effectiveness of deep features as a perceptual
  metric.
\newblock In {\em CVPR}, 2018.

\bibitem{KinD}
Yonghua Zhang, Jiawan Zhang, and Xiaojie Guo.
\newblock Kindling the darkness: A practical low-light image enhancer.
\newblock In {\em Proceedings of the 27th ACM International Conference on
  Multimedia}, MM '19, page 1632–1640, New York, NY, USA, 2019. Association
  for Computing Machinery.

\bibitem{RRDNet}
Anqi Zhu, Lin Zhang, Ying Shen, Yong Ma, Shengjie Zhao, and Yicong Zhou.
\newblock Zero-shot restoration of underexposed images via robust retinex
  decomposition.
\newblock In {\em 2020 IEEE International Conference on Multimedia and Expo
  (ICME)}, pages 1--6, 2020.

\bibitem{Zhu_Pan_Chen_Yang_2020}
Minfeng Zhu, Pingbo Pan, Wei Chen, and Yi Yang.
\newblock {EEMEFN}: Low-light image enhancement via edge-enhanced
  multi-exposure fusion network.
\newblock {\em Proceedings of the AAAI Conference on Artificial Intelligence},
  34(07):13106--13113, Apr. 2020.

\end{thebibliography}
}

\end{document}